\pdfoutput=1
\documentclass{jfm}
\usepackage{graphicx}
\usepackage{natbib}
\usepackage{upmath}
\usepackage{amssymb}
\usepackage{amsbsy}
\usepackage{amsmath}
\usepackage{graphics}
\usepackage{subfigure}
\usepackage{dcolumn}
\usepackage{bm}
\usepackage{hyperref}

\newcommand{\vx}{{\bf x}}

\newcommand{\bc}{\begin{center}}
\newcommand{\ec}{\end{center}}

\newcommand{\be}{\begin{equation}}
\newcommand{\ee}{\end{equation}}
\newcommand{\bea}{\begin{eqnarray}}
\newcommand{\eea}{\end{eqnarray}}

\title[]{Chaotic Mixing in Three Dimensional Porous Media}

\author[Daniel R. Lester, Marco Dentz, Tanguy Le Borgne]%
{Daniel R. Lester$^1$%
  \thanks{Email address for correspondence: daniel.lester@rmit.edu.au},\ns
Marco Dentz$^2$\break
and Tanguy Le Borgne$^3$}

\affiliation{$^1$School of Civil, Environmental and Chemical Engineering, RMIT University, 3000 Melbourne, Victoria, Australia\\[\affilskip]
$^2$Spanish National Research Council (IDAEA-CSIC), 08034 Barcelona, Spain\\[\affilskip]
$^3$Geosciences Rennes, UMR 6118, Universit\'{e} de Rennes 1, CNRS, 35042 Rennes, France}

\pubyear{2010}
\volume{650}
\pagerange{119--126}
\date{?; revised ?; accepted ?. - To be entered by editorial office}
\begin{document}

\maketitle

\begin{abstract}
Under steady flow conditions, the topological complexity inherent to all random 3D porous media imparts complicated flow and transport dynamics. It has been established that this complexity generates persistent chaotic advection via a three-dimensional (3D) fluid mechanical analogue of the baker's map which rapidly accelerates scalar mixing in the presence of molecular diffusion.  Hence pore-scale fluid mixing is governed by the interplay between chaotic advection, molecular diffusion and the broad (power-law) distribution of fluid particle travel times which arise from the non-slip condition at pore walls. To understand and quantify mixing in 3D porous media, we consider these processes in a model 3D open porous network and develop a novel stretching continuous time random walk (CTRW) which provides analytic estimates of pore-scale mixing which compare well with direct numerical simulations. We find that chaotic advection inherent to 3D porous media imparts scalar mixing which scales exponentially with longitudinal advection, whereas the topological constraints associated with 2D porous media limits mixing to scale algebraically. These results decipher the role of wide transit time distributions and complex topologies on porous media mixing dynamics, and provide the building blocks for macroscopic models of dilution and mixing which resolve these mechanisms.
\end{abstract}


\section{Introduction}

All porous media, whether random or ordered, granular or networked, heterogeneous or homogeneous,  are typified by the geometric and topological complexity of the pore-space~\citep{Scholz:2012,Vogel:2002}. This pore-space plays host to a wide range of fluid-borne processes including transport, mixing and dispersion, chemical reaction and microbiological activity, all of which are influenced by the flow structure and transport properties~\citep{Metcalfe:2012aa, Dentz::JCH::2011, deBarrosGRL2012, chiognaGRL2012}. Pore-scale fluid mixing plays a key role in the control of both fluid-fluid reactions (e.g. redox processes) and fluid-solid reactions (e.g. precipitation-dissolution processes), which are of importance for a range of subsurface operations, including CO$_2$ sequestration, contaminant remediation or geothermal dipoles management.  Whilst pore-scale flows are often smooth and steady (typically Stokesian or laminar),  the inherent topological complexity of the pore-space renders upscaling transport and mixing processes a challenging task.  

Because of their fundamental role in driving chemical reactions, mixing processes have received increasing attention in recent years
in the context of porous media flows ~\citep{Dentz::JCH::2011}. Two-dimensional laboratory experiments~\citep{Gramling2002, ATartakovskyWRR2008, deAnnaEST2014} and theoretical and modeling studies~\citep{Battiato:2009} have shown that upscaled chemical kinetics are not captured by classical macro-dispersion theories due to incomplete mixing at the pore scale. This points to a need for predictive theories for pore-scale concentration statistics which are couched in terms of the underlying medium properties. Lamellar mixing theories, developed in the context of turbulent flows \citep{Villermaux::Duplat::PRL::2003, Duplat::Villermaux::JFM::2008, Duplat::PF::2010}, have been applied and extended for the prediction of concentration
statistics in two-dimensional (2D) Darcy scale heterogeneous porous media \citep{LeBorgne2013, LB2015}. A central element of this theory is to quantify the link between fluid stretching and mixing.  In this context, linking the pore network topological properties to mixing dynamics is an essential step, which we explore in this study. 

While the topological constraints associated with the Poincar\'{e}-Bendixson theorem limit fluid stretching in two-dimensional (2D) steady flows to be algebraic, in three-dimensional (3D) steady flows much richer behaviour is possible. Indeed, the topological complexity inherent to all three dimensional random porous media has been shown to induce chaotic advection under steady flow conditions via a 3D fluid mechanical analogue of the baker's map~\citep{Lester:2013ab}. Such chaotic Lagrangian dynamics are well-known to rapidly accelerate diffusive mixing and scalar dissipation~\citep{Ottino:1989}, yet have received little attention with respect to pore-scale flow. From the perspective of transport dynamics, the distribution of pore sizes and shapes, together with no-slip boundary conditions at the pore walls, are known to impart non-Gaussian pore velocity distributions~\citep{Moroni2001B,  BijeljicEA:11, Kang:2014aa, Holzner2015}, which lead to a rich array of dispersion phenomena ranging from normal to super-diffusive. The continuous time random walk (CTRW) approach~\citep{Berkowitz2006} has been used to model this behaviours  \citep{BijeljicEA:2003, LeBorgneWRR2011, deanna13-prl, Kang:2014aa, Holzner2015} based on the transit time distributions over characteristic pore lengths, which reflect the distirbution of pore velocities.  The interplay of wide transit time distributions and chaotic advection at the pore-scale impacts both macroscopic transport and dispersion as well as pore-scale dilution.

Pore-scale chaotic advection has been shown~\cite{Lester:2014ab} to significantly suppress longitudinal dispersion arising from the no-slip wall condition due to transverse mixing generating an analogue of the Taylor-Aris mechanism. Conversely, the wide transit time distributions are expected to have a drastic impact on the dynamics of mixing in conjunction with chaotic advection as 
the transit times set the timescales over which significant stretching of a material fluid element occurs.  While the impact of broad transit time distributions on the spatial spreading of transported elements is well understood, their control on mixing dynamics is still an open question. As shown in~\citep{Lester:2014ab}, Lagrangian chaos generates ergodic particle trajectories at the pore-scale, and the associated decaying correlations allows the advection process to be modelled as a stochastic process. During advection through the pore-space, fluid elements undergo punctuated stretching and folding (transverse to the mean flow direction) events at stagnation points, leading to persistent chaotic advection in random porous media. Such dynamics have been captured in an idealized 3D random pore network model which comprises of a periodic network of uniform-sized pores which alternately branch and merge in the mean flow direction, leading to a high number density of stagnation points local to these pore junctions which generate fluid stretching. Whilst highly idealized, this network model contains basic features common to all porous media, namely topology-induced chaotic advection and no-slip boundary conditions, and so represents the minimum complexity inherent to all porous media.

In this paper, we develop a novel stretching CTRW which captures the advection and deformation of material elements through the porous matrix. Closure of this CTRW model in conjunction with a 1D advection-diffusion equation (ADE) describing diffusion transverse to highly striated, lamellar concentration distributions generated by pore-scale chaotic advection facilitates quantification of mixing and dispersion of scalar fields under the action of combined chaotic advection and molecular diffusion both within the pore space and across the macroscopic network. This formalism allows prediction of concentration PDF evolution within the pore-space and quantification of the impact of chaotic advection.

To simplify exposition, we first consider the somewhat artificial case of a steady state pore-scale mixing and dispersion of a concentration field which is heterogeneous at the pore-scale but homogeneous at the macro-scale continuously injected across all pores in a plane transverse to the mean flow direction. These results are then extended to the more realistic situation of the evolution of a solvent plume which is continuously injected as a point source. We compare these predictions for three distinctly different porous networks, a random 3D porous network which gives rise to Lagrangian chaos, an ordered 3D porous network which generates maximum fluid stretching, and an ordered 2D network which gives rise to non-chaotic dynamics. Hence the impact of both network topology and structure upon pore-scale mixing and dispersion is quantified. These results form a quantitative basis for upscaling of pore-scale dynamics to macroscopic mixing and transport models, and establish the impacts of ubiquitous chaotic mixing in 3D random porous media.

The paper is organized as follows; the mechanisms leading to topological mixing in 3D porous media are briefly reviewed in $\S$~\ref{sec:topomix}, followed by a description of the 3D open porous network model in $\S$~\ref{sec:3Dmodel}. Fluid stretching in this model network is considered in $\S$~\ref{Section:stretch}, from which a stretching CTRW model is derived in $\S$~\ref{sec:CTRW}. This model is then applied to quantify mixing in $\S$~\ref{sec:mixing}, and the implications for the evolution of the concentration PDF, mixing scale and dispersion are conisdered in $\S$~\ref{sec:diffmix} . The overall results are discussed in $\S$~\ref{sec:discussion} and finally conclusions are made in $\S$~\ref{sec:conclusions}.

\section{Topologically-Induced Fluid Deformation in 3D Porous Media}
\label{sec:topomix}

Topological complexity is a defining feature of all porous media - from
granular and packed media to fractured and open networks - these
materials are typified by a highly connected pore-space within which
the flow of continua arise. Such topological complexity is
characterised by the Euler characteristic $\chi$ (related to the
topological genus $g$ as $\chi=2(1-g)$) which measures the
connectivity of the pore-space~\citep{Vogel:2002} as
\begin{equation}
\chi=N-C+H,
\end{equation}
where $N$ is the number of pores, $C$ the number of redundant connections and $H$ the number of completely enclosed cavities. For porous media it is meaningful to consider the average number density of these quantities, where from computer tomography studies~\citep{Vogel:2002} it is found that typically $N$ is large whilst $C$, $H$ are small. Hence the number density of the Euler characteristic $\chi$ is uniformly found to be strongly negative~\citep{Scholz:2012,Vogel:2002}, reflecting the basic topological complexity which typifies all porous media. When a continuous fluid is advected through such media, a large number of stagnation points (non-degenerate equilibrium points) arise at the fluid/solid boundary as a direct result of this topological complexity. These stagnation points $\mathbf{x}_p$ are zeros of the skin friction vector field $\mathbf{u}(\mathbf{x})$ on the 2D boundary $\partial\mathcal{D}$ of the fluid domain $\mathcal{D}$, where $\mathbf{u}(\mathbf{x})$ may be defined as
\begin{equation}
\mathbf{u}(x_1,x_2,x_3):=\frac{\partial \mathbf{v}}{\partial x_3}.
\end{equation}
Here $\mathbf{v}(x_1, x_2, x_3)$ is the fluid velocity field, $x_3$ is the coordinate normal to the fluid boundary $\partial\mathcal{D}$ and $x_1$, $x_2$ are orthogonal coordinates tangent to this boundary. Whilst different definitions of the skin friction are possible~\citep{SuranaEA:06,Winkel/Bakker:88,Chong:2012}, these are all equivalent on the boundary $\partial\mathcal{D}$, as is the topology of the flow structure in the fluid domain $\mathcal{D}$. The Poincar\'{e}-Hopf theorem provides a direct relationship between the nature of the critical points $\mathbf{x}_p$ and the pore-space topology, such that the sum of the indices $\gamma_p$ of critical points $\mathbf{x}_p$ is related to the topological genus $g$ and Euler characteristic $\chi$ as
\begin{equation}
\sum_p \gamma_p(\mathbf{x}_p)=2(1-g)=\chi,\label{eqn:poincare_hopf}
\end{equation}
where the index $\gamma_p$ equals -1 for saddle-type zeros, +1 for node-type zeros and 0 for null zeros of the skin-friction field. Hence $|\chi|$ represents a lower bound for the number density of stagnation points under steady 3D Stokes flow, and these points impart significant fluid stretching into the local fluid domain. Digital imaging studies~\citep{Vogel:2002} measure the Euler characteristic across a broad range of porous media, from granular to networked, and find $\chi$ to be strongly negative, with number densities of the order $\sim$200-500 mm$^{-3}$. This large number density of stagnation points imparts a series of punctuated stretching events as the fluid continuum is advected through the pore-space. A relevant question is what role these stretching events play with respect to transport and mixing in porous media?

\begin{figure}
\begin{centering}
\begin{tabular}{c c c c}
\includegraphics[width=0.19\columnwidth]{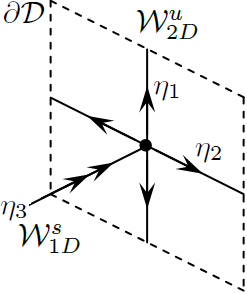}&
\includegraphics[width=0.19\columnwidth]{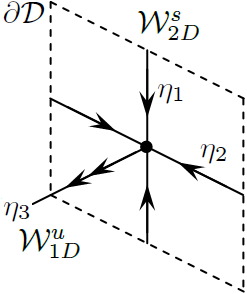}&
\includegraphics[width=0.26\columnwidth]{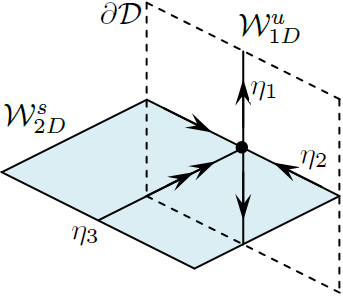}&
\includegraphics[width=0.26\columnwidth]{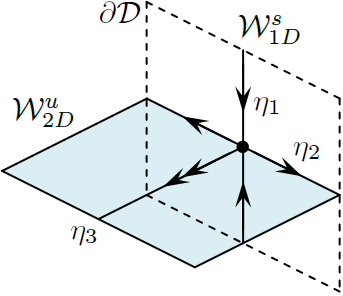}\\
I & II & III & IV
\end{tabular}
\end{centering}
\caption{Type I-IV non-degenerate equilibrium points (with stagnation points I,III and reattachment points II, IV) on the boundary $\partial\mathcal{D}$ and associated stable and unstable manifolds $\mathcal{W}^s$, $\mathcal{W}^u$. The double arrows reflect the sum $\eta_1+\eta_2+2\eta_3=0$.}\label{fig:dist}
\end{figure}

Stagnation points play a critical role with respect to the Lagrangian dynamics of 3D steady flows, as it is at these points that the formal analogy between transport in steady 3D volume-preserving flows and 1 degree-of-freedom Hamiltonian systems breaks down~\citep{Bajer_Moffat:1990,Bajer:1994} (such that the steady 3D dynamical system can no longer be expressed as an analogous unsteady 2D system), and such points are widely~\citep{Mezic_Wiggins:1994,Wiggins:2010} implicated in the creation of non-trivial Lagrangian dynamics. \citet{MacKay:1994,MacKay:2008} proposes that the stable $\mathcal{W}^s$ and unstable $\mathcal{W}^u$ manifolds which respectively correspond to the fluid contraction and stretching directions around stagnation points (shown in Figure~\ref{fig:dist}) form the ``skeleton'' of the flow, a set of surfaces of minimal transverse flux which organize transport within the fluid domain. If these manifolds which project into the fluid bulk are two-dimensional (hence co-dimension one), they form essentially impenetrable barriers which organise fluid transport and mixing.

Whilst the detailed dynamics of stagnation points and lines and their associated manifolds is more complicated~\citep{SuranaEA:06} than described herein, the main point is that 
the dimensionality of the stable or unstable manifolds which are normal to the suface $\partial\mathcal{D}$ is dictated by the topological index $\gamma_p$ or stagnation point type (saddle, node, null). Fluid deformation local to stagnation points $\mathbf{x}_p$ is quantified by the linearized skin friction tensor $\mathcal{A}$, the components $a_{ij}$ of which are given by the expansion
\begin{equation}
u_i=\sum_{j=1}^3a_{ij}(x_j-x_{p,j})+\mathcal{O}((x_j-x_{p,j})^2).
\end{equation}
Several workers~\citep{Winkel/Bakker:88,Chong:2012} have derived the normal form of the skin friction tensor $\mathcal{A}$ around a non-degenerate equilibrium point $\mathbf{x}_p$ in an incompressible flow as
\begin{equation}
\mathcal{A}=\left(
\begin{array}{ccc}
a_{11} & a_{12} & a_{13} \\
a_{21} & a_{22} & a_{23} \\
0 & 0 & -\frac{1}{2}(a_{11}+a_{22})
\end{array}
\right),
\label{eqn:skinfriction_tensor}
\end{equation}
where the eigenvalues $\eta$ of $\mathcal{A}$ satisfy $\eta_1+\eta_2+2\eta_3=0$, where $\eta_1$, $\eta_2$ are the eigenvalues in the skin friction boundary (with $\eta_1\leqslant\eta_2$), and $\eta_3$ is the interior eigenvalue. Hence $\eta_3>0$ for separation points and $\eta_3<0$ for reattachment points, and non-degenerate equilibrium points consist of one stable and two unstable eigenvalues or vice versa. As the linearisation (\ref{eqn:skinfriction_tensor}) indicates that the tangent boundary is an invariant plane, there exist four basic types of equilibrium points as shown in Figure~\ref{fig:dist}:

\begin{itemize}
\item type I separation point $\eta_3>0$, attractor for the skin friction field $\eta_1<0$, $\eta_2<0$,
\item type II reattachment point $\eta_3<0$, repeller for the skin friction field $\eta_1>0$, $\eta_2>0$,
\item type III separation point $\eta_3>0$, saddle for the skin friction field $\eta_1<0$, $\eta_2>0$,
\item type IV reattachment point $\eta_3<0$, saddle for the skin friction field $\eta_1<0$, $\eta_2>0$.
\end{itemize}

\begin{figure}
\begin{centering}
\includegraphics[width=0.85\columnwidth]{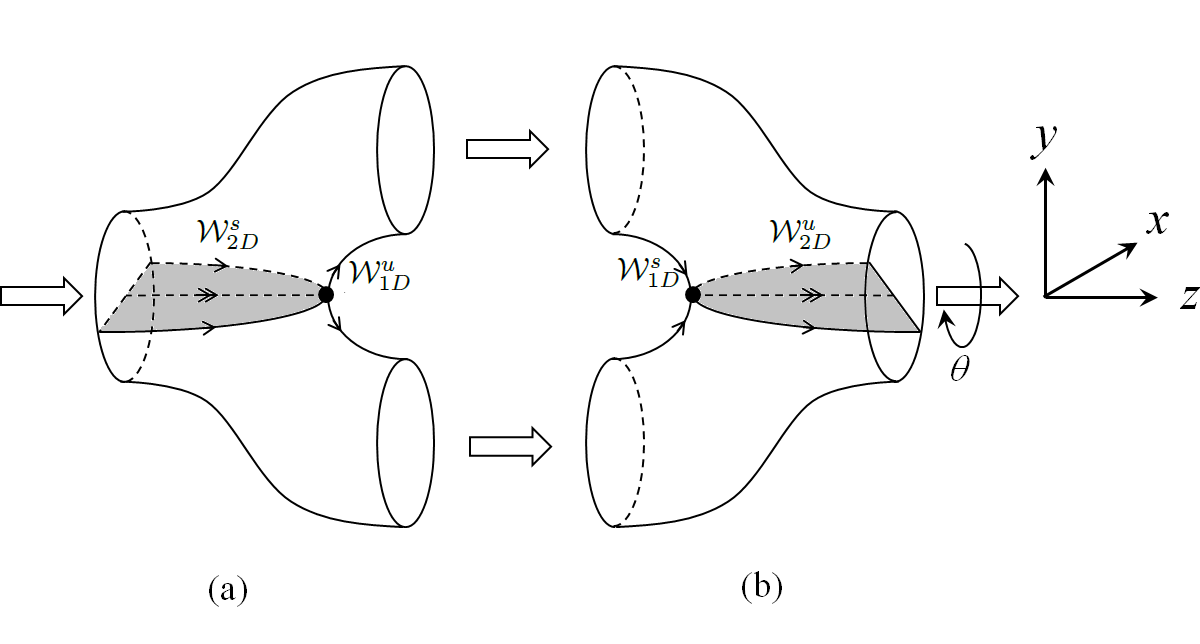}
\end{centering}
\caption{Schematic of pore branch (a) and pore merger (b) elements, with non-degenerate equilibrium stagnation (separation) points shown and associated 2D unstable (stable) manifolds, representing surfaces of locally minimum flux. Note the transverse orientation of angle $\theta$ of the minimum flux surfaces.}\label{fig:branch_topology}
\end{figure}

Type I and II points are of node type, whilst type III and IV points are saddle points (with $\eta_1<0$, $\eta_2>0$) where the respective stable and unstable invariant manifolds which propagate into the fluid interior are two-dimensional, and so form surfaces of locally minimal flux which act as barriers to transport. Due to the strongly negative Euler characteristic common to most porous media, then from (\ref{eqn:poincare_hopf}) such media admit a high number density of saddle-type points under steady 3D Stokes flow. Figure~\ref{fig:branch_topology} illustrates these concepts for a pore branch (a) and merger (b) (the union of which is termed a couplet) in an open porous network. Here, the reattachment (stagnation) point in (a) ((b)) must arise due to the basic topology of the pore branch (merger), and this point of saddle type III (IV) gives rise to a 2D manifold $\mathcal{W}^s_{2D}$ ($\mathcal{W}^u_{2D}$) which propagates into the fluid domain. It is important to note that whilst the geometry of pore branches and merges may vary significantly, the basic topology shown in Figure~\ref{fig:branch_topology} is universal to almost all porous media, whether porous networks or granular media.

The interaction of the 2D manifolds ($\mathcal{W}^s_{2D},\mathcal{W}^u_{2D}$) in the fluid bulk govern fluid transport and mixing. As is well-known from classical studies from Hamiltonian chaos, if two co-dimension 1 manifolds intersect transversely (via heteroclinic or homoclinic connections) then chaotic dynamics result, whereas smooth connections yield regular Lagrangian dynamics. For the pore junction shown in Figure~\ref{fig:branch_topology}, significant fluid stretching (compression) occurs transverse to the 2D manifolds (as indicated by the transverse 1D manifolds in Figure~\ref{fig:dist}), whilst folding of material elements occurs due to downstream advection local to the stagnation points. These actions are the constituent motions of the Smale horshoe map, a hallmark of chaotic dynamics in continuous systems. If the 2D manifolds $\mathcal{W}^s_{2D},\mathcal{W}^u_{2D}$ in Figure~\ref{fig:branch_topology} are oriented transversely (as indicated by the angle $\theta$), then persistent fluid stretching and folding can occur, whereas for symmetric connections ($\theta=0$), these deformations cancel and the symmetry of Lagrangian stretching histories leads to non-chaotic dynamics.

\begin{figure}
\begin{centering}
\includegraphics[width=0.9\columnwidth]{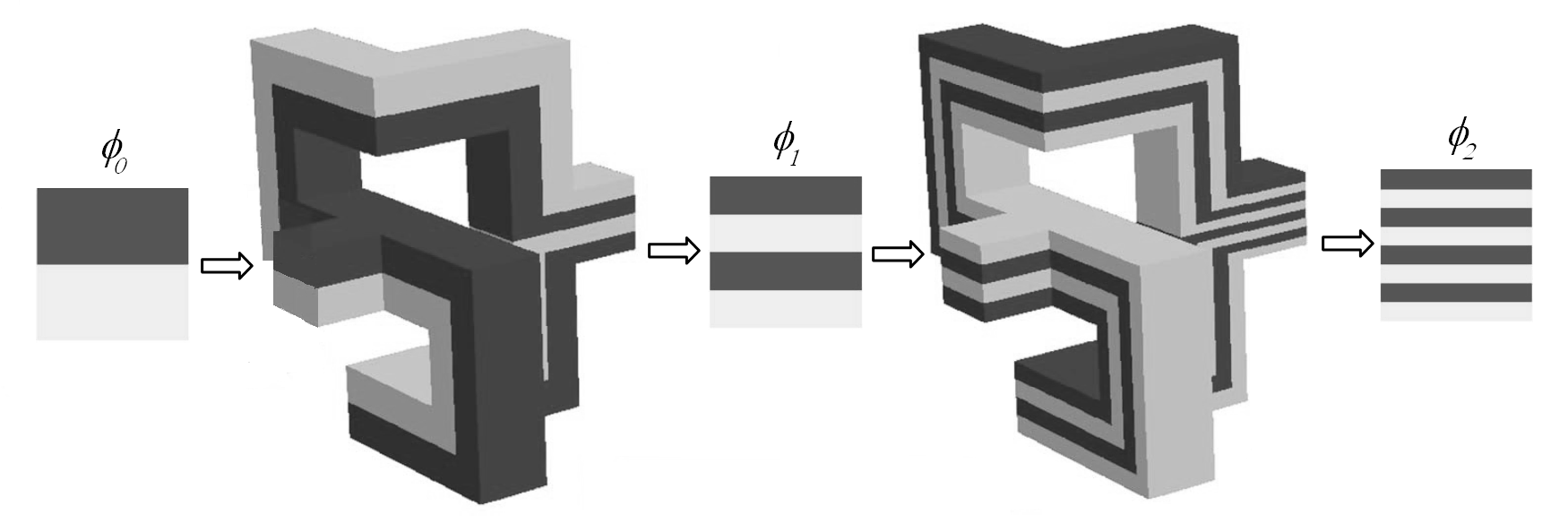}
\end{centering}
\caption{Schematic of the baker's \emph{flow}, a 3D fluid mechanical implementation of the baker's map arising from non-trivial pore branching and merging. Adapted from Carri\`{e}re~\citep{Carriere:2007}}\label{fig:3D_bakers}
\end{figure}

The dynamics of topologically complex systems is considered by~\citet{MacKay:2001}, who studied flow within a closed domain topologically equivalent to a connected pore branch and merger (Figure\ref{fig:branch_topology}) with the merger outlet also glued to the branch inlet, forming a closed domain of genus 2. As the topology of this closed domain is equivalent to that of the open domain in Figure~\ref{fig:branch_topology} with periodic boundaries, we make no distinction between the two systems. This closed flow is termed a baker's flow, which is a 3D fluid mechanical analogue of the baker's map, an archetypical chaotic map in Hamiltonian dynamics. This analogy corresponds to the case $\theta=\pi/2$, where transverse stretching and compression in the pore branch/merger leads to stretching and folding akin to the cutting and stacking of the Baker's map. \citet{MacKay:2001} shows that whilst this flow is not structurally stable, it is a volume-preserving analogue of the Lorentz system which is robustly mixing in the right parameter regime. This contention is supported by the numerical results of~\citet{Carriere:2007}, who considers 3D steady Stokes flow in a periodic duct with repeat branches and mergers as shown in Figure~\ref{fig:3D_bakers}. This flow is analogous to the pore couplet in Figure~\ref{fig:branch_topology} with $\theta=\pi/2$, and generates almost globally chaotic dynamics (with the exception of small KAM tori due to reentrant vortices in duct corners) in the transverse Poincar\'{e} section~\citep{Carriere:2007}. The measured Lyapunov exponent $\lambda_\infty=0.68$ is very close to the theoretical upper bound $\lambda_{\max} = \ln 2$ for steady 3D flows. Hence, the skeleton of the flow can generate chaotic mixing in porous networks quantitatively similar to strong theoretic mixing~\citep{Sturman:2008aa}.

These basic mechanisms persist in granular and packed media, where the minimal flux surface associated with type III stagnation points wrap downstream around a particle cluster, and likewise the minimal flux surface associated with a type IV stagnation point wraps upstream. Again the orientation angle $\theta$ between these surfaces plays a critical role with respect to the persistence of fluid deformation. The disordered nature of random porous media ensures that the transverse orientation $\theta\neq0$ condition is met with unit probability throughout the porous matrix, in which case the network structure imparts chaotic advection under steady flow conditions. Whilst porous media may exhibit other features such as surface roughness or pore tortuosity which may also generate chaotic advection~\citep{JonesEA:1989,OttinoWiggins:2004}, stagnation points are generic to topologically complex media, hence the mechanism described above represents a lower bound for chaotic dynamics inherent to porous media flow.

\section{3D Open Porous Network model}
\label{sec:3Dmodel}

\begin{figure}
\begin{centering}
\begin{tabular}{c c}
\includegraphics[width=0.5\columnwidth]{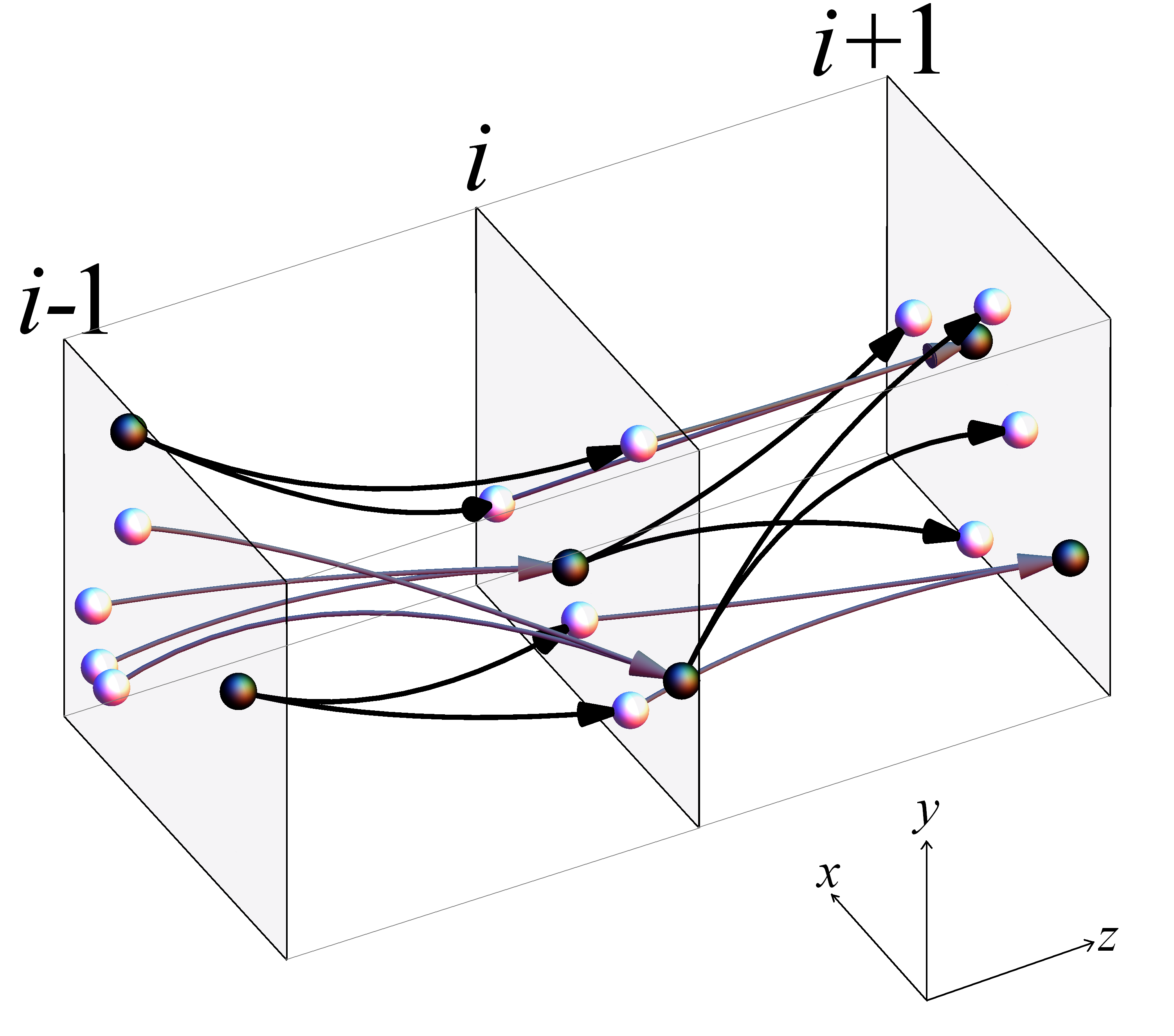}&
\includegraphics[width=0.5\columnwidth]{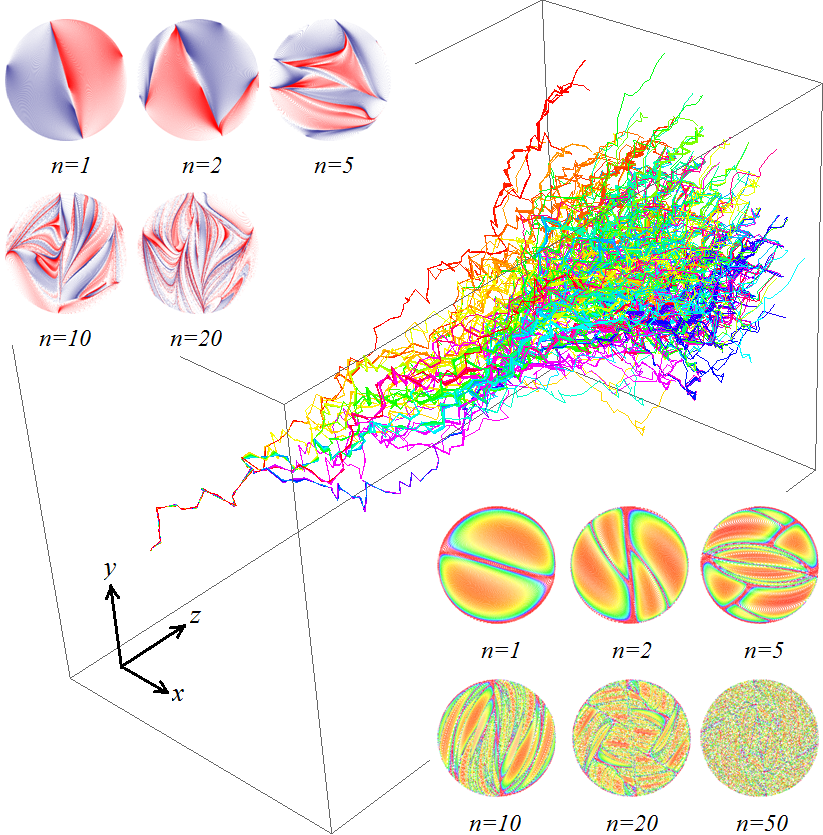}\\
(a) & (b)
\end{tabular}
\end{centering}
\caption{(a) Schematic of pore branches (black) and mergers (white) between mapping planes, with branch pores (black) and merge pores (white) shown. (b) (middle) Evolution of a typical non-diffusive dye plume in the pore network from continuous injection over a single inlet pore, (upper left) typical transverse distribution of non-diffusive coloured fluid particles with longitudinal pore number $n$, (lower right) typical distribution of residence times (pink = maximum, orange=minimum) with longitudinal pore number $n$.}\label{fig:pore_network}
\end{figure}

To study the impacts of such topology-induced chaotic advection upon transport and mixing in porous media, we consider a model 3D open porous network which consists of random network of connected pore junctions and mergers shown in Figure~\ref{fig:branch_topology}. We consider a non-trivial network of pore branches and mergers which is the simplest representation of an open porous network which may be considered homogeneous at the macroscale. To compose a random 3D network over the semi-infinite domain $\Omega:\mathbf{x}\in\mathbb{R}^3, z>0$, we use the pore branch and merge elements (shown in Figure~\ref{fig:branch_topology} (a), (b) respectively) to connect a series of so-called ``mapping planes'' oriented parallel to the $(x,y)$ plane, distributed along the $z$-axis at integer multiples of the pore element length $\ell$. Each mapping plane consists of an infinite number of reflections in the $x$, $y$ directions of a periodic unit cell ${x,y}\in[0,1]\times[0,1]$ which contains $3N$ randomly-located non-overlapping pores, as shown in Figure~\ref{fig:pore_network}. The location of the $j$-th pore in the $i$-th mapping plane is labelled $\mathbf{r}_{i,j}=(x_{i,j},y_{i,j},i \ell)$.

Of these $3N$ pores, $N$ are randomly labelled as ``branch'' pores (i.e. pores which branch next in the positive $z$ direction), and the remaining $2N$ pores are labelled as ``merge pores'' (i.e. those which are about to merge). A branch pore at plane $i$ is connected to two merge pores at plane $i+1$ by a pore branch element located between these planes, and conversely two branch pores at plane $i$ are connected with a single merge pore at plane $i$ by a pore merge element as per Figure~\ref{fig:branch_topology}. Connections between merge and branch pores in adjacent planes are made by identifying unique nearest neighbour groupings of a single merge pore in one plane and two branches in the adjacent plane (accounting for periodicity in the $(x,y)$-plane), such that all pores undergo sequential branching and merging as they propagate along the $z$-coordinate. Connections are restricted such that a pair of branch pores at plane $i$ common to a single merge pore at $i-1$ do not share the same merge pore at $i+1$. This restriction both eliminates ``degenerate'' pore branch/merger couplings and ensures the orientation of connections is essentially random.

To accommodate connections between the merge pore locations $\mathbf{r}_{m_1},\mathbf{r}_{m_2}$ and the branch pore location $\mathbf{r}_{b_1}$ at adjacent mapping planes, the pore branch and merge elements are rotated and stretched such that the $\hat{\mathbf{e}}_y$, $\hat{\mathbf{e}}_z$ vectors respectively are aligned along $\mathbf{r}_{m_1}-\mathbf{r}_{m_2}$, $\mathbf{r}_{b_1}-\frac{1}{2}(\mathbf{r}_{m_1}+\mathbf{r}_{m_2})$, and the element height and length (as oriented in Figure~\ref{fig:branch_topology}) respectively are re-scaled to $||\mathbf{r}_{m_1}-\mathbf{r}_{m_2}||$, $||\mathbf{r}_{b_1}-\frac{1}{2}(\mathbf{r}_{m_1}+\mathbf{r}_{m_2})||$. The orientation angle $\theta$ in the $(x,y)$ plane of a pore branch/merge element is then

\begin{equation}
\theta=\arctan\left[\frac{(\mathbf{r}_{m_1}-\mathbf{r}_{m_2})\cdot\hat{\mathbf{e}}_y}{(\mathbf{r}_{m_1}-\mathbf{r}_{m_2})\cdot\hat{\mathbf{e}}_x}\right].\label{eqn:twistangle}
\end{equation}

Hence the set $S$ of random pore locations $\mathbf{r}_{i,j}$ for $i=0:\infty$, $j=1:3N$ completely defines a realization of the 3D open porous network, and the set of all realizations $\mathcal{S}$ of the porous network form a ensemble which is ergodic and stationary. Due to the elimination of degenerate pore branch/merger couplings, the distribution of orientation angles $\theta$ are uncorrelated across and within mapping planes, and so transport within the 3D open porous network may be described by a multidimensional Markov process in space.

To model transport within the porous network, we consider fluid advection via the velocity field $\mathbf{v}$ subject to Stokes flow $\mu\nabla^2\mathbf{v}+\nabla p=0$ within the pore branch element $\Omega$ driven by a differential pressure $p|_{z=0}-p|_{z=L}=\Delta p$, subject to no-slip boundary conditions $\mathbf{v}\cdot\mathbf{n}|_{\partial\Omega}=0$.  The flow field is calculated numerically to order $10^{-16}$ RMS accuracy using the finite-volume CFD package ANSYS-CFX with double precision calculations. The advection dynamics for passive tracer particles within the branch element $\Omega$ under Stokes flow can be represented by the spatial $\mathcal{M}^*$ and temporal $\mathcal{T}^*$ maps
\begin{align}
&\mathcal{M}^*:\{x_{r,i},y_{r,i}\}\mapsto\{x_{r,i+1},y_{r,i+1}\},\\
&\mathcal{T}^*:t_{i}\mapsto t_{i+1}
\end{align}
where $\{x_{r,i},y_{r,i}\}$ are the $x,y$ particle positions relative to the inlet pore of $\Omega$ (where $x_{r,i}^2+y_{r,i}^2=0,1$ respectively corresponds to the pore centre and boundary), and $\{x_{r,i+1},y_{r,i+1}\}$ are the $x,y$ particle positions relative to the outlet pore of $\Omega$. Note that $x_{r,i+1},y_{r,i+1}$ are independent of the specific outlet pore the particle travels to. Similarly, the temporal map $\mathcal{T}^*$ describes the residence time from the inlet to outlet planes. From the CFD results, the exact maps $\mathcal{M}^*$, $\mathcal{T}^*$ are remarkably well approximated (within error $\epsilon\sim 10^{-3}$) by the simple analytic maps
\begin{align}
&\mathcal{M}:\{x_r,y_r\}\mapsto
\begin{cases}
&\{x_r,2y_r-\sqrt{1-x_r^2}\}\quad \text{if} \quad y_r>0\\
&\{x_r,2y_r+\sqrt{1-x_r^2}\}\quad \text{if} \quad y_r\leqslant0
\end{cases},\label{eqn:Mmap}\\
&\mathcal{T}:t\mapsto t+\frac{1}{1-x_r^2-y_r^2}.\label{eqn:Tmap}
\end{align}
These approximate maps greatly simplify description of the advection dynamics in the pore branch element $\Omega$ and preserve the essential features of advective transport, namely stretching of fluid elements by a factor of 2 in the $xy$-plane to preserve cross-sectional area, and the Poiseuille distribution of $\mathcal{T}$ (scaled such that the minimum advection time is unity) reflects the no-slip boundary conditions. Transport in a pore merge element is quantified by the inverse spatial map $\mathcal{M}^{-1}$, and the spatial residence time distribution is given by the composition $\mathcal{T}\circ \mathcal{M}^{-1}$. Although the advective maps $\mathcal{M}$, $\mathcal{M}^{-1}$ are not area-preserving, $\mathcal{M}$ and $\mathcal{M}^{-1}$ are implicitly volume-preserving when the fluid velocity in the $z$-direction is accounted for. Concatenation of a pore branch and merger results in area preservation and zero net fluid deformation, as reflected $\mathcal{M}^{-1}\circ \mathcal{M}=I$, where $I$ is the identity operator.

Pore branches and mergers at arbitrary relative orientation $\theta$ in the $xy$-plane break this symmetry and do not result in such degeneracy. Rather, the rotation operator
\begin{equation}
R(\theta):\{x,y\}\mapsto\{x\cos\theta+y\sin\theta,y\cos\theta-x\sin\theta\},\label{eqn:rotation}
\end{equation}
coupled with the pore branch $\mathcal{M}$ and merge $\mathcal{M}^{-1}$ maps define transport in pore branches and mergers at arbitrary angles
\begin{align}
&\mathcal{S}(\theta)=R(\theta)\circ \mathcal{M}\circ R(-\theta),\\
&\mathcal{S}^{-1}(\theta)=R(\theta)\circ \mathcal{M}^{-1}\circ R(-\theta),
\end{align}
where $\mathcal{S}(\theta_1)\circ \mathcal{S}(\theta_2)\neq I$ for $\theta_1\neq \theta_2$. Coupled with the residence time distribution $\mathcal{T}$, the composite maps $\mathcal{S}$, $\mathcal{S}^{-1}$  quantify advective transport through the 3D porous network, and the position and residence time of fluid particles can be rapidly propagated via these approximate maps. The distribution of alignment angles $\theta$ dictates fluid stretching within the network, where the sequence of angles $\theta_1,\theta_2,\dots$ for a particular trajectory can impart either chaotic or non-chaotic advection~\citep{Lester:2013ab}, and the associated infinite-time Lyapunov exponent spans $\lambda_\infty\in[0,\ln 2]$. Conversely, random media with $\theta$ uniformly distributed as a Markov process generates globally chaotic dynamics which are typically weaker than that of ordered media.

This basic pore network model has been used~\citep{Lester:2013ab,Lester:2014ab} to study the impact of Lagrangian chaos upon dispersion in porous media, i.e. the spatial spreading of transport particles. Here, we investigate its impact on mixing properties, i.e. the distribution and temporal dynamics of concentration statistics by explicitly coupling fluid deformation with molecular diffusion. Figure~\ref{fig:pore_network} (b) illustrates the macroscopic evolution of a dye trace simulation propagated by the composite maps $\mathcal{S}$, $\mathcal{S}^{-1}$ following a single pore through a realization of the random network. The lower right sub-figure in Figure~\ref{fig:pore_network} (b) illustrates the creation of fine-scale structure via chaotic advection within the pore-scale (from an initially segregated distribution of half blue/half red points across all pores), and the upper left subfigure shows evolution of the residence time distribution evolved via the composite temporal maps $\mathcal{S}\circ\mathcal{T}$, $\mathcal{S}^{-1}\circ\mathcal{T}$. Note there is no evidence of folding of fluid elements in the $xy$-plane as folding occurs around stagnation points as fluid elements are advected downstream in the $z$-direction.

\section{Stretching and Compression in 3D Porous Networks\label{Section:stretch}}

\subsection{Stretching and Compression in 3D Ordered Networks}

To calculate transport, mixing and dispersion with the random 3D porous network model, we first consider the deformation of a continuously injected 2D material filament under a mean flow in the $z$-direction as it propagates over pore branches and mergers through the network. To determine the evolution of this 2D material filament throughout in 3D random network, it shall prove convenient to consider this steady 2D filament as an evolving 1D material line in the mapping-planes, the constituent points of which are projected via the advective maps $\mathcal{S}(\theta)$, $\mathcal{S}(\theta)^{-1}$. 

Whilst fluid stretching does occur in the $z$-direction (due to both deformation in the parabolic flow field and iterated stretching and compression over pore branches and mergers), the impact upon mixing is considered negligible due to alignment of the continuously injected 2D filament with the mean flow direction. Hence the gross impact of chaotic advection upon dispersion and mixing under steady-state conditions is to accelerate mixing transverse to the mean flow direction $z$. As such, only the stretching dynamics in the $xy$-plane are required to quantify mixing under steady flow conditions.

As the advective map $\mathcal{M}$ represents the flow (in the dynamical systems sense) of fluid particles from the inlet to outlet pores, the 2D fluid deformation gradient tensor $\mathbf{F}_\text{2D}$ over a pore branch is given by the gradient which may be linearised as
\begin{equation}
\mathbf{F}_{\text{2D}}=\frac{\partial\mathcal{M}_i}{\partial x_j}=
\left(
\begin{array}{cc}
1 & \pm \frac{x}{\sqrt{1-x^2}}  \\
0 & 2 
\end{array}
\right)
\approx
\left(
\begin{array}{cc}
1 & 0 \\
0 & 2
\end{array}
\right),
\end{equation}
and likewise fluid deformation over a pore merger is given by $\mathbf{F}_{\text{2D}}^{-1}$. CFD simulations of Stokes flow through the branch element $\Omega$ shows that this linearisation well approximates fluid deformation in the branch and merge elements. The gross action of the pore branch is to stretch fluid elements by a factor of 2 in the $y$-direction, and simultaneously contract elements by a factor of $\frac{1}{2}$ in the $z$-direction (not reflected in $\mathbf{F}_{\text{2D}}$, whilst conversely the pore merger contracts fluid elements by a factor of $\frac{1}{2}$ in the $y$-direction and stretches by a factor of $2$ in the $z$-direction. Similar to the advective maps $\mathcal{M}$, $\mathcal{M}^{-1}$, the concatenated branch and merge deformation tensors generate zero net stretching ($\mathbf{F}_{\text{2D}}\cdot\mathbf{F}^{-1}_{\text{2D}}=\mathbf{1}$), but when reoriented, concatenation of these tensors can generate persistent fluid stretching.

To quantify stretching and compression in an ordered 3D porous network, we consider the reoriented deformation tensors
\begin{subequations}
\begin{align}
&\mathbf{S}_s=\mathbf{R}(\theta_s)\cdot\mathbf{F}_{\text{2D}}\cdot\mathbf{R}^{-1}(\theta_s),\label{eqn:linestretch}\\
&\mathbf{S}_c=\mathbf{R}(\theta_c)\cdot\mathbf{F}^{-1}_{\text{2D}}\cdot\mathbf{R}^{-1}(\theta_c),\label{eqn:linecompress}
\end{align}
\end{subequations}
where $\mathbf{R}(\theta)$ is the rotation matrix associated with reorientation through angle $\theta$ in the $xy$-plane. Fluid deformation over a coupled pore branch/merge element (couplet) is then
\begin{equation}
\begin{split}
\mathbf{S}&=\mathbf{S}_s\cdot\mathbf{S}_c,\\
&=\mathbf{R}(\theta_s)\cdot\mathbf{F}_{\text{2D}}\cdot\mathbf{R}(\Delta)\cdot\mathbf{F}_{\text{2D}}^{-1}\cdot\mathbf{R}^{-1}(\Delta)\cdot\mathbf{R}^{-1}(\theta_s),\\
&=\mathbf{R}(\theta_s)\cdot\mathbf{D}(\Delta)\cdot\mathbf{R}^{-1}(\theta_s),\label{eqn:pore_stretch}
\end{split}
\end{equation}
where $\Delta:=\theta_s-\theta_c$. The logarithm of the eigenvalues of $\mathbf{S}$, $\mathbf{D}(\Delta)$ are the stretching rates over a couplet
\begin{equation}
\lambda_{b,m}=\pm\ln(\zeta+\sqrt{\zeta^2-1}),
\end{equation}
where $\zeta=\frac{9}{8}-{1}{8}\cos 2\Delta$. For pore branches and mergers which are parallel ($\Delta=n\pi, n=0,1,2,\dots$), there is no net stretching ($\zeta=0$), whereas orthogonally oriented elements ($\Delta=\pi/2+n\pi, n=0,1,2,\dots$) generate maximum stretching $\lambda_c=\ln 2$ akin to the baker's map. Net fluid deformation over a series of $n$ concatenated couplets is then given by the series
\begin{equation}
\begin{split}
\boldsymbol\Lambda_n&=\mathbf{R}(\theta_n)\cdot\mathbf{D}(\Delta_n)\cdot\mathbf{R}^{-1}(\theta_n)\cdot\dots\cdot\mathbf{R}(\theta_1)\cdot\mathbf{D}(\Delta_1)\cdot\mathbf{R}^{-1}(\theta_1),\\
&=\mathbf{R}(\theta_n)\cdot\left(\prod_{j=1}^n\mathbf{L}(\delta_j,\Delta_j)\right)\cdot\mathbf{R}^{-1}(\theta_n),
\end{split}
\end{equation}
where $\mathbf{L}(\delta,\Delta):=\mathbf{R}(\delta)\cdot\mathbf{D}(\Delta)$, $\delta_j=\theta_{j+1}-\theta_j$ for $j=1:n-1$ and $\delta_n=\theta_1-\theta_n$. As such, the deformation tensor $\mathbf{L}(\delta,\Delta)$ quantifies fluid deformation over a series of connected pore couplets, and net fluid deformation depends strongly upon the series of angles $\Delta_j$ within couplets and the series of relative orientation $\delta_j$ between couplets.

\begin{figure}
\begin{centering}
\begin{tabular}{c c}
\includegraphics[width=0.5\columnwidth]{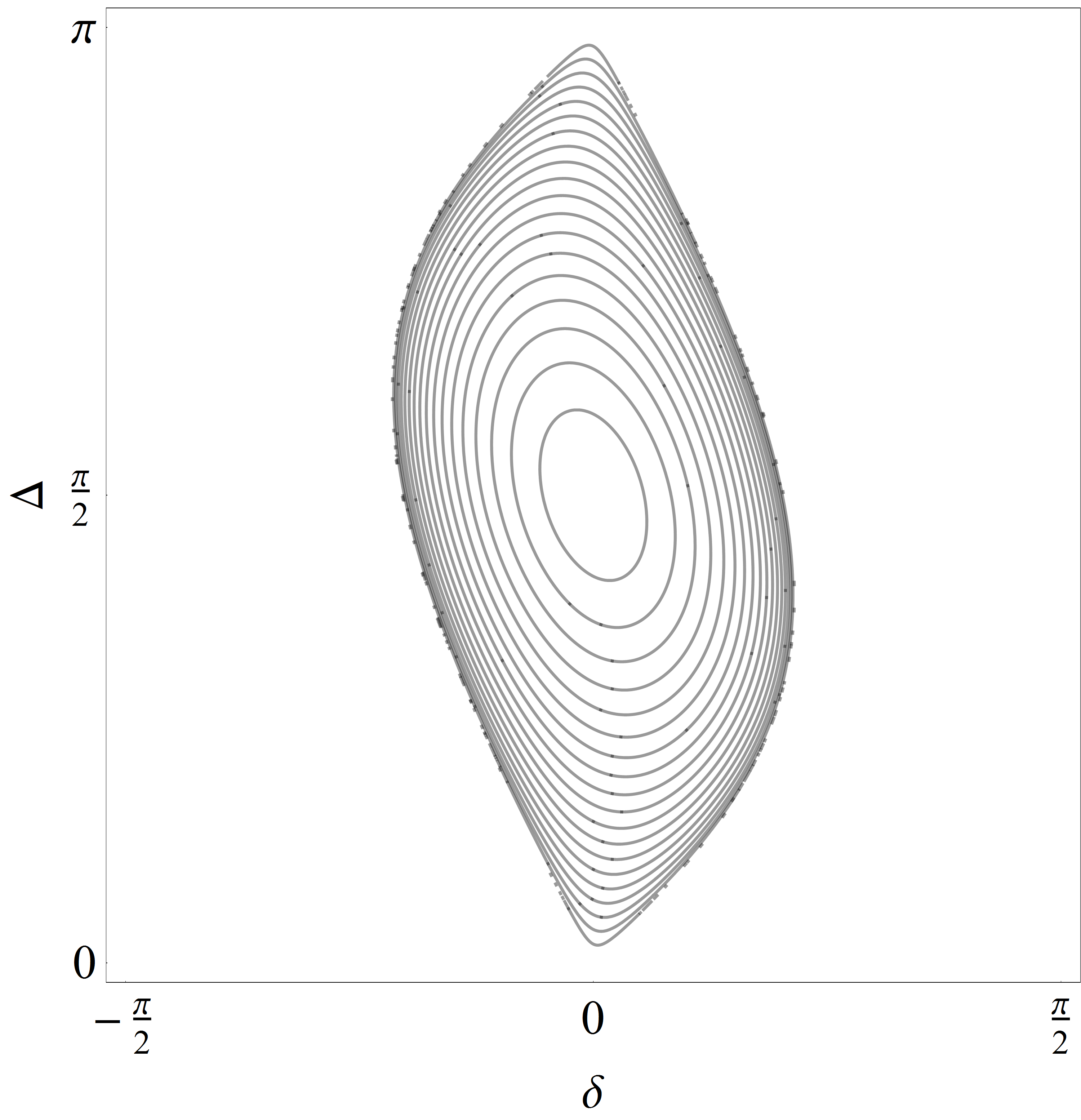}&
\includegraphics[width=0.5\columnwidth]{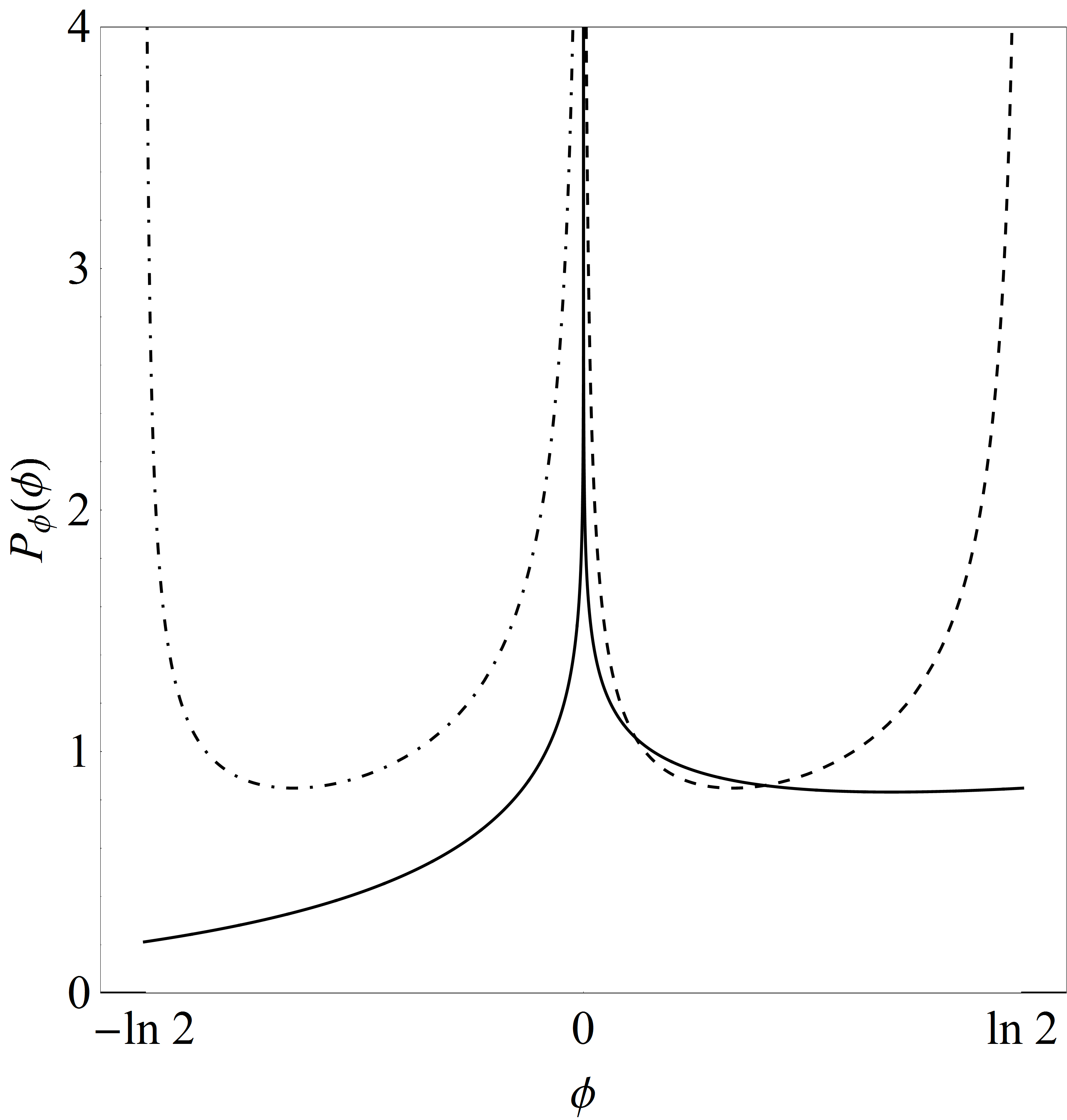}\\
(a) & (b)
\end{tabular}
\end{centering}
\caption{(a) Contour plot of the distribution of logarithmic stretching rates $\lambda(\delta,\Delta)$ in ordered 3D networks as a function of the orientation angles $\delta$, $\Delta$, between and within pore couplets. The stretching rate $\lambda$ varies between $\lambda=0$ for $|\zeta|\leqslant1$ and the theoretical maximum for continuous systems $\lambda=\ln 2$ for $\zeta=\frac{5}{4}$ at $\Delta=\pi/2$, $\delta=0$. (b) PDF of logarithmic stretching rates $P_\phi(\phi)$ (solid), $P_{\phi_s}(\phi_s)$ (dashed) and $P_{\phi_c}(\phi_c)$ (dot-dashed) in random 3D networks. Note $P_\phi(\phi)$ is strongly peaked but asymmetric around $\phi=0$.}\label{fig:stretching rates}
\end{figure}

For ordered media which consists of a fixed $\delta$ and $\Delta$, net fluid deformation is quantified by the infinite-time Lyapunov exponent $\lambda$ given by the logarithm of the eigenvalues of the deformation tensor $\mathbf{L}(\delta,\Delta)$
\begin{equation}
\lambda(\delta,\Delta)=\pm\ln|\zeta+\sqrt{\zeta^2-1}|,
\end{equation}
where $\zeta=\frac{9}{8}\cos\delta-\frac{1}{8}\cos\left(2\Delta+\delta\right)$. As shown in Figure~\ref{fig:stretching rates} (a), zero net deformation occurs for $|\zeta|\leqslant 1$, whilst maximum deformation ($\lambda=\ln 2$) occurs for $\Delta=(n+\frac{1}{2})\pi$, $\delta=n \pi$. Note that zero net stretching is possible even if pore branches and mergers are oriented transversely ($\Delta=(n+\frac{1}{2})\pi$); whilst maximum fluid stretching occurs within a couplet, if a neighbouring couplets are oriented transversely ($\delta=(n+\frac{1}{2})\pi$), then the transverse stretching in the next couplet cancels the net deformation. Hence ordered 3D porous networks represent extreme cases with respect to fluid stretching and deformation; whilst a large class ($|\zeta|\leqslant 1$) of ordered media do not exhibit chaotic advection ($\lambda=0$), certain networks exhibit the maximum theoretic stretching ($\lambda=\ln 2$) for continuous systems. 

The stretching rate $\lambda(\delta,\Delta)$ quantifies stretching of the 2D material filament as it propagates through the 3D network in the that the length $l$ of the filament cross section in the $xy$-plane grows with longitudinal pore number $n$ or distance $z$ as
\begin{equation}
l(z)=l_0\exp\left(\lambda(\delta,\Delta) n\right)=l_0\exp\left(\lambda (\delta,\Delta)\frac{z}{\ell}\right),
\end{equation}
where $l_0$ is the initial length of the filament cross section. In the presence of chaotic advection where $\lambda>0$, unbounded growth of the filament cross section is accommodated by striated packing within pores as per Figure~\ref{fig:pore_network}.

\subsection{Stretching and Compression in 3D Random
  Networks}

In contrast to ordered 3D networks, the orientation angle $\theta$ between pore elements in the model random 3D network follows a Markov process, and so the average stretching rate $\lambda_\infty$ (Lyapunov exponent) is
\begin{equation}
\lambda_\infty=\frac{1}{\pi^2}\int_{-\pi/2}^{\pi/2} d\delta \int_{0}^\pi d\Delta\,\lambda(\delta,\Delta)\approx 0.1178,
\end{equation} 
which closely matches the numerically computed asymptotic stretching rate calculated from the advective maps $\mathcal{S}(\theta)$, $\mathcal{S}(\theta)$ for random $\theta$~\citep{Lester:2013ab}.

We consider evolution of an infinitesimal element within the 1D material line represented by the vector $\mathbf{l}$, where $\mathbf{l}$ evolves with pore number $n$ over pore branch and merge elements respectively as
\begin{subequations}
\begin{align}
\mathbf{l}_{2n+1}& =
\mathbf{R}(\theta_s)\cdot\mathbf{F}_{\text{2D}}\cdot
\mathbf{R}^{-1}(\theta_s)\cdot\mathbf{l}_{2n},\label{eqn:2Dlinestretch}\\
\mathbf{l}_{2n+2}&=
\mathbf{R}(\theta_c)\cdot\mathbf{F}^{-1}_{\text{2D}}\cdot
\mathbf{R}^{-1}(\theta_c)\cdot\mathbf{l}_{2n+1},\label{eqn:2Dlinecompress}
\end{align}
\end{subequations}
where $\theta_s$, $\theta_c$ are the orientation angles of the pore branch, merger with
respect to the $y$-direction as per Figure~\ref{fig:branch_topology}. The length $l=|\mathbf{l}|$ of the line element then evolves via the two-step process 
\begin{subequations}
\label{eqn:rhoincrements}
\begin{align}
l_{n+1} &= l_n \rho_s(\varphi_n),\\
l_{n+2} &= l_{n+1} \rho_c(\varphi_{n+1}),
\end{align}
\end{subequations}
where $\varphi_n$ are the orientation angles of the pore branch or merger with respect to the line element and $\rho_s$, $\rho_c$ respectively are the relative elongation due to stretching and compression within a pore branch and merger:
\begin{subequations}\label{eqn:rho_sc}
\begin{align}
\rho_s(\varphi_s)&=\sqrt{1+3\cos^2\varphi_s},\label{eqn:rho_s}\\
\rho_c(\varphi_c)&=\frac{1}{2}\sqrt{4-3\cos^2\varphi_c}.\label{eqn:rho_c}
\end{align}
\end{subequations}
For random media, the pore branches and mergers are randomly oriented such that $\varphi_s$, $\varphi_c$ are uniformly distributed over $\theta\in[-\pi,\pi]$. As such,
it is not necessary to consider the orientation of the material line
as $\varphi_s$, $\varphi_c$ are also uniformly distributed over
$[-\pi,\pi]$, and so the mean relative stretching $\lambda_s$ and compression
$\lambda_c$ within a pore branch and merger are
\begin{subequations}
\begin{align}
\label{lambdas}
\lambda_s&=\frac{1}{2\pi}\int_{-\pi}^{\pi}d\varphi_s\ln[\rho_s(\varphi_s)]\approx 0.405465,\\
\label{lambdac}
\lambda_c&=\frac{1}{2\pi}\int_{-\pi}^{\pi}d\varphi_c\ln[\rho_c(\varphi_c)]\approx -0.287682,
\end{align}
\end{subequations}
where the sum of these stretching rates recovers the infinite-time
Lyapunov exponent
$\lambda_s+\lambda_c=\lambda_\infty\approx0.11783$. The full distribution of stretching
rates can also be derived from (\ref{eqn:rho_sc}) as
\begin{align}
P_{\rho_s}(\rho_s)&=\frac{4}{\pi\sqrt{\rho_s^2-1}\sqrt{4-\rho_s^2}},\quad\rho_s\in[1,2],\label{eqn:rhos_distn}\\
P_{\rho_c}(\rho_c)&=\frac{4}{\pi\sqrt{1-\rho_c^2}\sqrt{4\rho_c^2-1}},\quad\rho_c\in[\frac{1}{2},1],\label{eqn:rhoc_distn}
\end{align}
and the total stretch $\rho:=\rho_s\rho_c$ over a couplet is
distributed as
\begin{equation}
\begin{split}
P_{\rho}(\rho)=&\int_{\frac{1}{2}}^1 d\rho_c' \int_1^{2} d\rho_s' P_{\rho_s}(\rho_s')P_{\rho_c}(\rho_c')\delta(\rho-\rho_c'\rho_s'),\\
=&\frac{4}{\pi^2}\left|\frac{\rho}{\rho^2-1}\right|K\left(1-\frac{9\rho^2}{4\rho^2-4}\right),\rho\in[\frac{1}{2},2],
\end{split}
\end{equation}
where $K$ is the complete elliptic integral of the first kind. To
derive the total stretching over many pore couplets, it is convenient
to consider evolution of the logarithmic length $s:=\ln l$ of a line
element, where $s$ evolves over a couplet via the one-step additive process 
\begin{equation}
\label{stretchprocess}
s_{n+2}=s_n+\phi_s+\phi_c=s_n+\phi, 
\end{equation}
where $\phi_s=\ln \rho_s$, $\phi_c=\ln \rho_c$, and
$\phi=\phi_s+\phi_c=\ln \rho$ are distributed as
\begin{align}
&P_\phi(\phi)=\exp(\phi)P_\rho(\exp\phi),\,\,\,\,\phi\in[-\ln 2,\ln 2],\label{eqn:phi_stretch}\\
&P_{\phi_s}(\phi_s)=\exp(\phi_s)P_{\rho_s}(\exp\phi_s),\,\,\,\,\phi_s\in[0,\ln 2],\label{eqn:phi_s_stretch}\\
&P_{\phi_c}(\phi_c)=\exp(\phi_c)P_{\rho_c}(\exp\phi_c),\,\,\,\,\phi_c\in[-\ln 2,0]\label{eqn:phi_c_stretch},
\end{align}
%
Asymmetry of the stretching PDF
$P_\phi(\phi)$ for random $\varphi_s$, $\varphi_c$ leads to persistent
fluid stretching as shown in Figure~\ref{fig:stretching rates}(b) and reflected by the  Lyapunov exponent $\lambda_\infty$. This asymmetry arises from the asymmetry
between the stretching and compression processes; within a pore branch
stretching is enhanced as the material line rotates toward the maximum
stretching direction, whilst compression is retarded in a pore merger
due to rotation away from the contraction direction. Hence whilst
transverse fluid stretching and compression are equally partitioned in random 3D networks, the asymmetry between stretching and compression generates persistent chaotic advection.

As both the mean ($\lambda_\infty$) and the variance ($\sigma^2\approx
0.11436$) of $P_\phi(\phi)$ are bounded, then the sum of the
stretching increments $\phi$ converges with $n$ toward a Gaussian
distribution via the central limit theorem, such that $s_n$ is approximately
distributed as
\begin{equation}
P_{s_n}(s)\approx\frac{1}{\sigma\sqrt{n\pi}}
\exp\left[-\frac{(s-s_0-\frac{n}{2}
    \lambda_\infty)^2}{n\sigma^2}\right], \label{eqn:s_gaussian}
\end{equation}
where the initial element length $s_0=\ln l_0$. Thus, the PDF
of the non-dimensional strip elongation $\rho_n = l_n / l_0$ is given by the
lognormal distribution
\begin{align}
\hat p_\rho(\rho,n) = \frac{1}{\rho \sqrt{\pi n \sigma^2}} \exp\left[-
  \frac{(\ln \rho - n \lambda_\infty/2)^2}{n \sigma^2} \right],
\label{eq:pdf_rho:nw}
\end{align}
which captures convergence of the distribution of finite Lyapunov
exponents $\lambda_n=s_n/n$ toward $\lambda_\infty$ as shown in \citep{Lester:2013ab}.

As 2D material filaments flow through pore branches and mergers in random 3D porous networks, they undergo a series of punctuated stretching and compression events around stagnation points which leads to net exponential stretching transverse to the mean flow direction. The length of material elements transverse to the mean flow direction are distributed log-normally, and the mean and variance respectively grow with $n$ as
$n\lambda_\infty/2$ and $n\sigma^2/2$.

Notice that the PDF~\eqref{eq:pdf_rho:nw} quantifies the point-wise
elongation statistics. When sampling the lamella elongation in space,
however, the sampling rate is proportional to the length of the
lamella. This is of particular importance because concentration statistics of the heterogeneous mixture are determined by spatial sampling across the lamellae. The elongation PDF weighted by the lamellae length, $p_\rho(\rho,n) \propto \rho \hat p_\rho(\rho,n)$, is
given by 
\begin{align}
p_\rho(\rho,n) = \frac{1}{\rho \sqrt{2 \pi \sigma_{\ln \rho}^2}} \exp\left[-
  \frac{(\ln \rho - \mu_{\ln \rho})^2}{2 \sigma_{\ln \rho}^2} \right], 
\label{eq:pdf_rho}
\end{align}
where we defined 
\begin{align}
\mu_{\ln \rho} =n (\lambda_\infty + \sigma^2)/2, && \sigma^2_{\ln
  \rho} = \sigma^2 n /2. 
\end{align}
For simplicity of notation in the following we set $
\Lambda_\infty = \lambda_\infty + \sigma^2$. 
To study the impact of pore-scale chaotic dynamics on fluid mixing and
transport, we compare several different types of open networks, from
random 3D networks ($\lambda=\lambda_\infty$), ordered 3D networks
with maximum stretching ($\lambda=\ln 2$), and 2D networks
($\lambda=0$) which have the same dynamics as ordered 3D networks with
zero stretching.
\section{Stretching Continuous Time Random Walk}
\label{sec:CTRW}

The stretching dynamics for ordered and random media developed above
provide inputs for the evolution of mixing in porous networks. We propose to describe the deformation process via a stretching Continuous Time Random Walk (CTRW), whereby fluid elements undergo a series of
punctuated stretching and folding events as they propagate through the
topologically complex pore network. CTRW models have been previously
developed for modeling dispersion processes, based on the observation
that Lagrangian velocities in porous media tend to be non-Markovian in
time but Markovian in space \citep{Le-Borgne:2008aa, deanna13-prl}.
This property is a consequence of the nature of the considered velocity fluctuations,
which are created by solid structures that can often be described
by a characteristic length scale (e.g. typical grain size or
permeability field correlation length). As low
Lagrangian velocities are maintained over this
characteristic length scale, they are likely persist over large time scales. 
Spatial Markov models formalized in the CTRW framework capture this broad
transit time distribution \citep{Le-Borgne:2008ab, BijeljicEA:11}. 
Similarly to advective motions, fluid stretching events are likely to persist over a finite
correlation scale, which is the pore length in the present study, and
will therefore occur in broadly distributed random times.
This is the basis for the proposed stretching CTRW model.

As fluid mixing involves the interplay of advection and diffusion, and 
the diffusion process is dependent upon the advection time over pore branches and mergers, 
it is necessary to extend the two-step stretching process (\ref{eqn:rhoincrements}) 
to include quantification of advection as a two-step CTRW:
\begin{subequations}\label{eqn:twostepCTRW}
\begin{align}
l_{2n+1} &= l_{2n} \rho_s(\varphi_n),&& t_{2n+1}=t_{2n}+\Delta t_{2n},\\
l_{2n+2} &= l_{2n+1} \rho_c(\varphi_{2n+1}),&& t_{2n+2}=t_{2n+1}+\Delta t_{2n+1}, 
\end{align}
\end{subequations}
where the time increment $\Delta t_n$ represents the advection time between pore elements. 
The stretching CTRW~\eqref{eqn:twostepCTRW} consists of punctated random stretching $\rho_s$ and compression $\rho_c$ processes which occur within each pore. The duration of the stretching and compression events is distributed as $\psi(\Delta t)$. Thus, the stretching CTRW quantifies the elongation of a material segment at subsequent downstream
positions $z_n$ through $l_n$, and the deformation time through the
temporal random walk $t_n$. The deformation rates during a stretching
or compression transition are constant and given by 
\begin{align}
\gamma_n = \frac{\ln[\rho_s(\varphi_n)]}{\Delta t_n}, && \gamma_{n+1} =
\frac{\ln[\rho_c(\varphi_{n+1})]}{\Delta t_{n+1}}, 
\end{align}
and so, we obtain for the strip length $l(t)$ 
\begin{align}
l(t) = l_{n_t} \exp[\gamma_{n_t} (t - t_{n_t})], 
\end{align}
where the renewal process $n_t = \sup(n|t_n \leq t)$ measures the
number of steps needed to arrive at time $t$. 

From the temporal map $\mathcal{T}$, the advection of fluid particles
through a pore branch or merger is well-approximated by the Poiseuille
flow
\begin{align}
v(r) = v_0 \left(1 - \frac{r^2}{R} \right)
\end{align}
with $v_0$ the maximum velocity and $R$ the pore radius.
Thus, in the absence of diffusion, the transition time over a pore
branch element at radius $r$ is given by $\Delta t(r) =
\ell/v(r)$ with $\ell$ the length of the pore
branch element. Due to the ergodicity of chaotic orbits, particles
sample the whole cross-section with equal probability, and so the
transition time distribution is given by the Pareto distribution
\begin{align}
\label{psi}
\psi(\Delta t) = \frac{1}{\Delta t_a}
\left(\frac{\Delta t}{\Delta t_a}\right)^{-2},  &&
\Delta t  \geq \Delta t_a.
\end{align}
where $\Delta t_a = \ell / v_0$. For porous media which are more complex than the
model pore network under consideration here, pore velocities are often
found to align according to exponential or stretched exponential
distributions~\cite[][]{Moroni2001B, Kang:2014aa, Siena2014, Holzner2015}. The corresponding transit times shows a similar long time behavior as~\eqref{psi}.  

Derivation of the distribution of average stretch $\rho$ over a couplet allows the two step CTRW (\ref{eqn:twostepCTRW}) to be replaced with the one-step CTRW
\begin{align}
\label{onestepCTRW}
s_{2n+2}=s_{2n}+\phi_{2n}, &&
t_{n+2}=t_{2n}+\Delta t_{2n}+\Delta t_{2n+1}.
\end{align}
Note that the stretching rate for this one-step CTRW is not uniform over a couplet (given by the time increment $\Delta t_{2n}+\Delta t_{2n+1}$), but rather is highly oscillatory due to the mean stretching rate $\lambda_s$, $\lambda_c$ over a pore branch or merger respectively. Whilst such oscillations do not impact the overall stretching rate $\lambda_\infty$, they may have significant implications for molecular diffusion as oscillatory stretching can leave a diffusive signature over the couplet. We shall return to this issue in $\S$\ref{sec:mixing}. 

\begin{figure}
\begin{centering}
\begin{tabular}{c c}
\includegraphics[width=0.5\columnwidth]{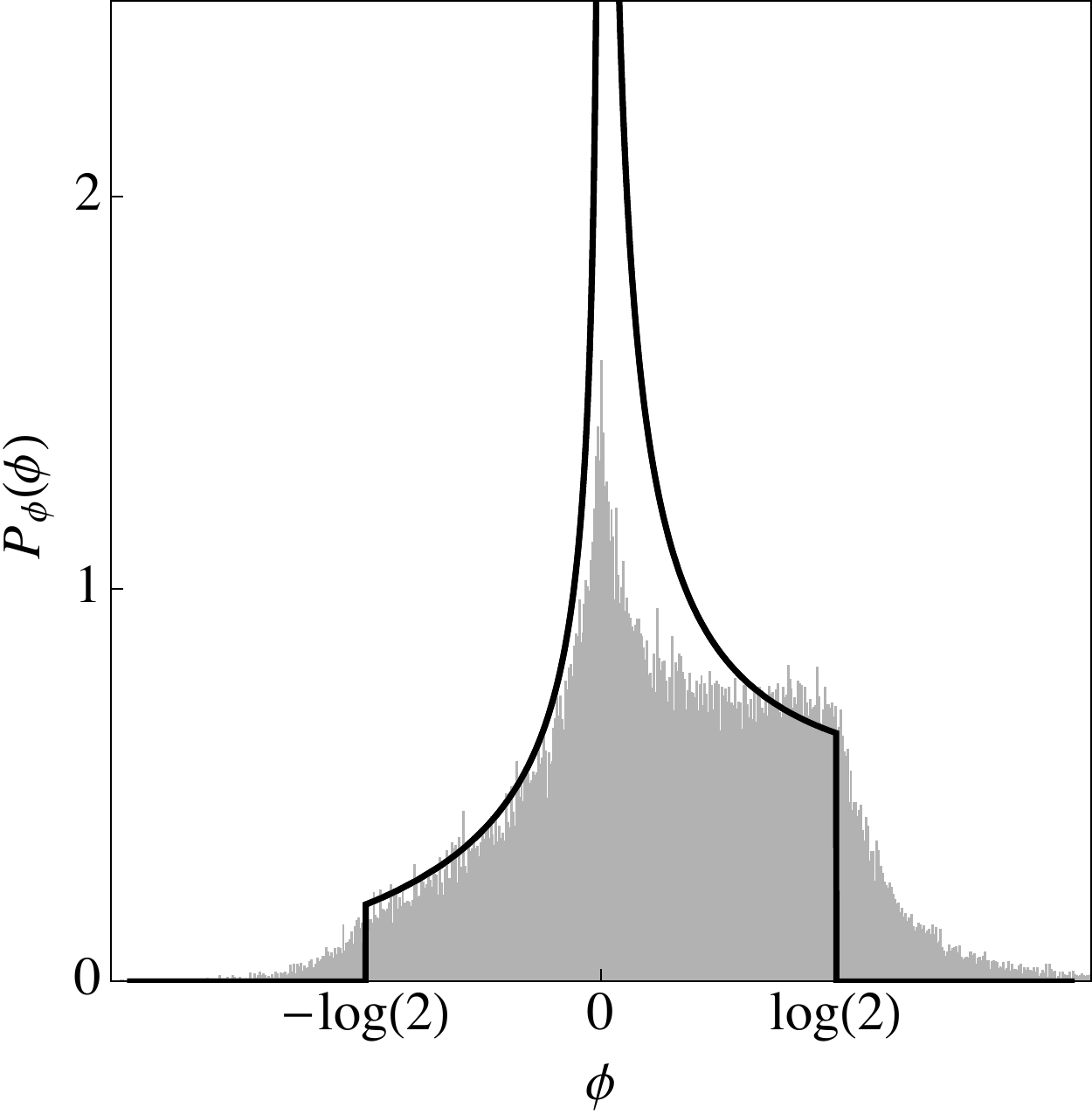}&
\includegraphics[width=0.5\columnwidth]{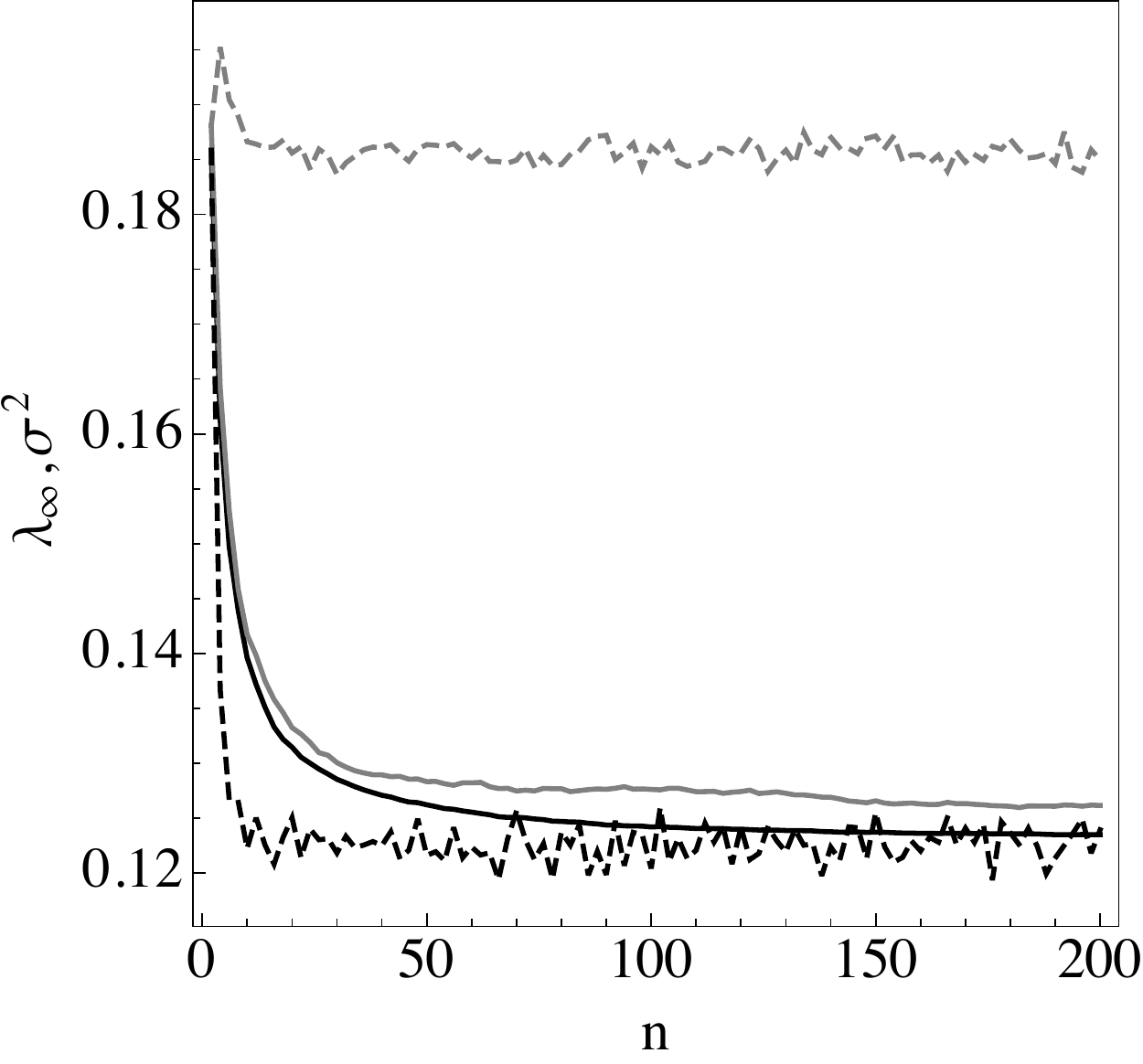}\\
(a) & (b)
\end{tabular}
\end{centering}
\caption{(a) Comparison of the PDFs of logarithmic stretching  $P_\phi(\phi)$ over a couplet computed from the nonlinear advective map  $\mathcal{M}$ (solid gray) or derived the linearised deformation tensor $\mathbf{F}_{\text{2D}}$ (black line). (b) Evolution of the mean  $\lambda_\infty$ (black, solid) and variance $\sigma^2$ (gray,  solid) of material deformation with the number of pores $n$,  and  convergence of the increment of the mean $\langle  \phi_{n+1}\rangle - \langle \phi_{n} \rangle$ (black, dashed) and variance $\langle  (\phi_{n+1}-\langle \phi_{n+1} \rangle)^2 \rangle - \langle  (\phi_{n}-\langle \phi_{n} \rangle)^2  \rangle$ (gray, dashed), respectively reflecting anti-correlation and non-stationarity of the adjective map $\mathcal{M}$.}\label{fig:nonlinear}
\end{figure}

Whilst the length-weighted distribution of material lengths $\rho_n=l_0\exp{s_n}$ given by the PDF (\ref{eq:pdf_rho}) characterizes material deformation over a couplet under action of the 2D deformation tensors $\mathbf{F}_{\text{2D}}$,$\mathbf{F}_{\text{2D}}^{-1}$, these operators do not capture the full deformation due to linearisation of the map $\mathcal{M}$. The nonlinear shear deformation in $\mathcal{M}$ generates additional deformation as reflected by the comparison of stretching rates over a couplet between
the advective map $\mathcal{M}$ and linearised deformation tensor $\mathbf{F}_{\text{2D}}$ in Figure~\ref{fig:nonlinear}(a). Deformation due to nonlinear shear extends the range of stretching beyond the bounds $\rho\in[1/2,2]$ and increases the asymmetry of the stretching process, increasing both the variance and mean of the stretching distribution over a couplet to $\sigma^2\approx 0.1277$ $\lambda_\infty\approx 0.124$  respectively.

\begin{figure}
\begin{centering}
\includegraphics[width=0.7\columnwidth]{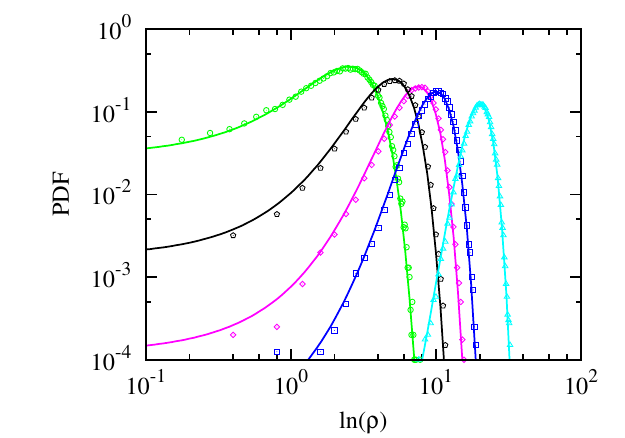}
\caption{Comparison of the lognormal form~\eqref{eq:pdf_rho} of the
  PDF of $\ln \rho$ (lines) to the PDFs obtained numerically from the one-step CTRW
  model~(\ref{onestepCTRW}) for (circles) $n = 20$, (pentagons) $n = 40$,
  (diamonds) n = 60, (squares) $n = 80$, (triangles) $n = 160$.} \label{fig:rhopdf}
\end{centering}
\end{figure}

Furthermore, the stretching process over a couplet under $\mathcal{M}$ is both non-stationary and anti-correlated due to preferential alignment of initially random oriented material elements with the pore
boundary as they undergo stretching. This is purely a geometric effect in that highly striated fluid elements must preferentially align tangentially with the pore boundary to allow packing within a finite domain. As the stretching process under $\mathcal{M}$ rapidly converges (after ~20 pores) to a stationary process, the increment of the mean $\langle  \phi_{n+1} \rangle - \langle \phi_{n} \rangle$ and variance $\langle  (\phi_{n+1}-\langle \phi_{n+1} \rangle)^2 \rangle - \langle (\phi_{n}-\langle \phi_{n} \rangle)^2  \rangle$ quickly converge as
per Figure~\ref{fig:nonlinear}(b). Conversely, anti-correlation causes these moments to converge more slowly toward the asymptotic values $\lambda_\infty\approx 0.12770$, $\sigma^{2}\approx 0.12366$. Rather
than develop a stochastic model which fully captures non-stationarity and anti-correlation, we approximate this process by using the uncorrelated and stationary one-step stretching CTRW (\ref{onestepCTRW}), where the increment $\phi$ is from the distribution shown in Figure~\ref{fig:nonlinear}(a) with mean and variance given by these asymptotic values. Figure~\ref{fig:rhopdf} shows that the lognormal form~\eqref{eq:pdf_rho} parameterized by the asymptotic values of $\lambda_\infty\approx 0.12770$, $\sigma^{2}\approx 0.12366$ compares very well with PDFs of $\ln \rho$ from direct numerical calculation.
%


\section{Scalar Mixing and Fluid Stretching CTRW}\label{sec:mixing}

To describe the interplay of fluid stretching and molecular diffusion in the generation of scalar mixing in the pore space we employ a lamellar mixing model~\cite[][]{Ranz79,Duplat::Villermaux::JFM::2008,LB2015}, where fluid stretching is quantified by the CTRW model developed in $\S$\ref{sec:CTRW}. Under exponential fluid stretching, an evolving concentration field conforms to a highly striated lamellar structure, where these lamellae align with the dominant stretching direction(s) and are simultaneously stretched longitudinally and compressed transversely.

We define a single lamella in the 3D pore space as the 2D invariant manifold which arises from the continuous injection of solute into a single pore in the inlet plane ($z=n=0$). For simplicity we consider the injected solute concentration to be distributed as a concentration strip of length $l_0$ with a transverse Gaussian profile of maximum concentration $c_0$ and variance $\sigma_0^2$, and the 2D lamella sheet (a material surface) evolves via the downstream advection of the 1D ridge of the injected concentration distribution. Hence the 2D lamella sheet acts as a backbone for the longitudinally evolving concentration field $c$, and this 2D surface is deformed and stretched as it evolves under advection throughout the pore-space, branching into multiple pores (as per Figure~\ref{fig:pore_network}) but remaining smooth and continuous.

The ensemble of all 2D lamellae arising from injection across all injection pores are homogeneously distributed throughout the pore network and so lamellae from different injection pores coexist within a given pore (as per Figure~\ref{fig:pore_network}) with spacing that decays as the exponential stretching rate in the longitudinal direction, but do not intersect due to their invariant nature. Conversely, the 3D concentration distributions associated with 2D lamellae may overlap, resulting in the eventual coalescence of lamellae in the longitudinal direction.

The steady 3D concentration field $c(\mathbf{x})$ local to an individual 2D lamella sheet arises from the interplay of fluid deformation and molecular diffusion local to this structure, and satisfies the steady advection diffusion equation (ADE)
\begin{equation}
-\mathbf{v}(\mathbf{x})\cdot\nabla c(\mathbf{x})+D_m\nabla^2 c(\mathbf{x})=0,\label{eqn:3D_ADE}
\end{equation}
subject to the zero-flux condition $\nabla c(\mathbf{x})\cdot\mathbf{n}|_{\partial\mathcal{D}}=0$, where $\mathbf{v}(\mathbf{x})$ is the 3D velocity field, $D_m$ molecular diffusivity, and $\mathbf{n}$ is the outward normal from the fluid domain $\mathcal{D}$. Under the assumption that the spatial concentration gradients are small in the $z$-direction (corresponding to minimal curvature of the lamella in this direction), we ignore longitudinal diffusion and pose the steady 3D ADE (\ref{eqn:3D_ADE}) as an equivalent unsteady 2D ADE in the $xy$-plane
\begin{equation}
\frac{\partial c}{\partial t}=v_z\frac{\partial c}{\partial z}=-\mathbf{v}_\bot(x,y,t)\cdot\nabla_{\bot}c(x,y,t)+D_m\nabla^2_\bot c(x,y,t),\label{eqn:sep_ADE}
\end{equation}
where $v_z$ is the $z$-component of $\mathbf{v}(\mathbf{x})$, $\mathbf{v}_\bot$ and $\nabla_\bot$ denote the $xy$-components of velocity and gradient operator respectively. The $z$-coordinate is also parameterised in terms of the Lagrangian travel time $t=t(z)$ along a given fluid particle trajectory as
\begin{equation}
t(z)=\int_0^z dz^\prime \frac{1}{v_z(z^\prime)}.\label{eqn:L_time}
\end{equation}
Whilst the travel time $t(z_0)$ for fixed $z=z_0$ also varies over the  $xy$-plane, (\ref{eqn:sep_ADE}) provides a convenient basis for solution of the concentration field based upon a CTRW for the advection time and deformation history of fluid elements.

The lamellar mixing model is based upon posing (\ref{eqn:sep_ADE}) in terms of the material coordinates $\{\eta,\zeta\}$ in the $xy$-plane, where $\zeta$ is the coordinate along the lamella, and $\eta$ is the transverse coordinate, hence $c(x,y,t)\mapsto c(\eta,\zeta,t)$. Under exponential fluid stretching, concentration gradients along the lamella ($\partial_\zeta$) decay exponentially whilst the gradients normal to the lamella ($\partial_\eta$) are maintained, and so (\ref{eqn:sep_ADE}) simplifies to the 1D lamellar transport equation~\cite[][]{Ranz79, Duplat::Villermaux::JFM::2008, Villermaux_12, LeBorgne2013}
\begin{align}
\label{eqn:zetaADE}
\frac{\partial}{\partial t}c(\eta,\zeta,t)= \gamma(t,\zeta) \eta \frac{\partial}{\partial \eta}c(\eta,\zeta,t) + D_m \frac{\partial^2}{\partial \eta^2}c(\eta,\zeta,t),
\end{align}
where the stretching rate $\gamma(t,\zeta)\equiv\partial_\zeta v_\zeta=-\partial_\eta v_\eta$ arises from a first order expansion of the $\eta$-component of the velocity field as
\begin{equation}
\begin{split}
v_\eta(\eta,\zeta,t)&=v_\eta(0,\zeta,t)+\eta\frac{\partial}{\partial \eta}v_\eta(\eta,\zeta,t)+\dots,\\
&=v_\eta(0,\zeta,t)-\gamma(t,\zeta)\eta+\dots,
\end{split}
\end{equation}
under the change of coordinates to the material frame. The stretching rate $\gamma(t,\zeta)$ can be expressed in terms of the lamella elongation in the $xy$-plane $l(t,\zeta)$ as 
\begin{align}
\gamma(t,\zeta)\equiv\frac{\partial v_\zeta}{\partial \zeta}=\frac{d \ln l(t,\zeta)}{dt}.
\end{align}
%

Note that the 1D nature of (\ref{eqn:zetaADE}) renders it only valid up to the coalescence of lamellae in the pore-space. Whilst methods \citep{Duplat::Villermaux::JFM::2008, Villermaux_12, LeBorgne2013} exist to propagate the concentration field beyond coalescence and are readily applicable to this problem, they are beyond the scope of this present study. As such, (\ref{ADEstrip}) describes the evolution of the concentration field $c(\eta,\zeta,t)$ with Lagrangian time $t$ associated with either a single 2D lamella sheet or multiple lamellae via superposition.

As a lamella injected at a given pore at the injeciton plane $z=n=0$ is distributed over many pores as it is advected downstream, segments of the material coordinate $\zeta$ are likewise distributed over multiple pores. Whilst in principle this may complicate solution of the evolving concentration field, for the homogeneous injection protocol all pores at the injection plane $z=n=0$ are seeded with the same initial condition, and so the re-distribution of lamellar segments between pores does not impact the concentration distribution at a statistical level. Hence the ADE (\ref{eqn:zetaADE}) describes the spatial concentration distribution of a series of lamellar segments in an arbitrary pore, irrespective of origin.  Whilst this simplification no longer holds for inhomogeneous injection protocols such as point or line sources, results for the homogeneous injection protocol may be readily extended as demonstrated in $\S$\ref{sec:diffmix}.

Due to negligible transport in the $\zeta$ direction, (\ref{eqn:zetaADE}) may also be posed with respect to a single material fluid 
element trajectory that represents an infinitesimal lamella element, and so describes evolution of the associated transverse concentration $c(\eta,t)$ profile as
\begin{align}
\label{ADEstrip}
\frac{\partial}{\partial t}c(\eta,t)= \gamma(t) \eta \frac{\partial}{\partial \eta}c(\eta,t) + D_m \frac{\partial^2}{\partial \eta^2}c(\eta,t).
\end{align}
This particle-based description of fluid mixing is now compatible with the CTRW stretching framework, which describes the advection and deformation of fluid elements through the pore network, in terms of the advection time and deformation history. As such, we do not seek to directly solve the evolution of an entire lamella, but rather solve the evolving concentration profile $c(\eta,t_n)$ over a representative ensemble of points which recovers the same statistics of the ensemble of all lamellae in the pore-space due to ergodicity.

If we consider fluid advection and stretching as stochastic processes, then for given $z_n$ or $n$ the advection time $t_n = t(z_n)$ is distributed randomly according to the waiting time distribution (\ref{psi}). Hence $c(\eta,t_n)$ is a random variable through its dependence on both the random advection time $t_n$ and random deformation $\gamma_n = \gamma(t_n)$. Under the assumption of independent lamellae, the PDF of concentration at a cross-section at $z_n$ is obtained by (i) sampling the concentration values across the lamella, and (ii) sampling between the lamellae. This interpretation allows the lamellar mixing equation (\ref{ADEstrip}) to describe evolution of the transverse concentration profile with advection time and  stretching history of any part of any lamellae in the the entire 3D porous network.

In order to quantify scalar mixing, we first solve for the concentration profile $c(\eta,t)$ across a lamella. The initial concentration distribution across the strip is assumed to follow the Gaussian profile
\begin{equation}
\label{c0}
c(\eta,0) = \frac{c_0}{\sqrt{2\pi \sigma_0^2}}\exp\left(-\frac{\eta^2}{(2 \sigma_0^2)}\right),
\end{equation}
such that the maximum initial concentration
$c_m(0) = c_0/\sqrt{2\pi\sigma_0^2}$ scales inversely with the
initial variance $\sigma_0^2$, and the initial strip length is denoted $l_0$. For simplicity of exposition we first consider the case where this initial length and concentration profile is uniform across all pores at the injection plane $z=0$ across the entire network, and so the concentration distribution at a given longitudinal pore number $n$ is uniform across all transverse pores. We then extend these results to the case of a continuously injected point-source plume (as shown in Figure~\ref{fig:pore_network}(b)) in $\S$\ref{sec:diffmix}. 

\begin{figure}
\begin{centering}
\begin{tabular}{c c c}
\includegraphics[width=0.33\textwidth]{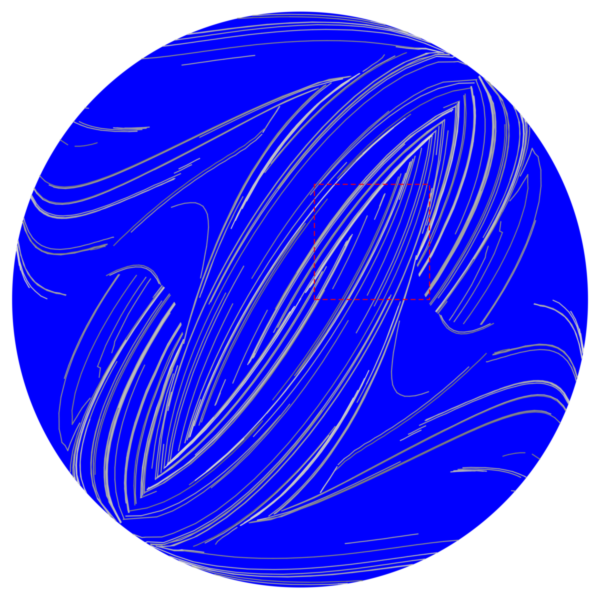}&
\includegraphics[width=0.33\textwidth]{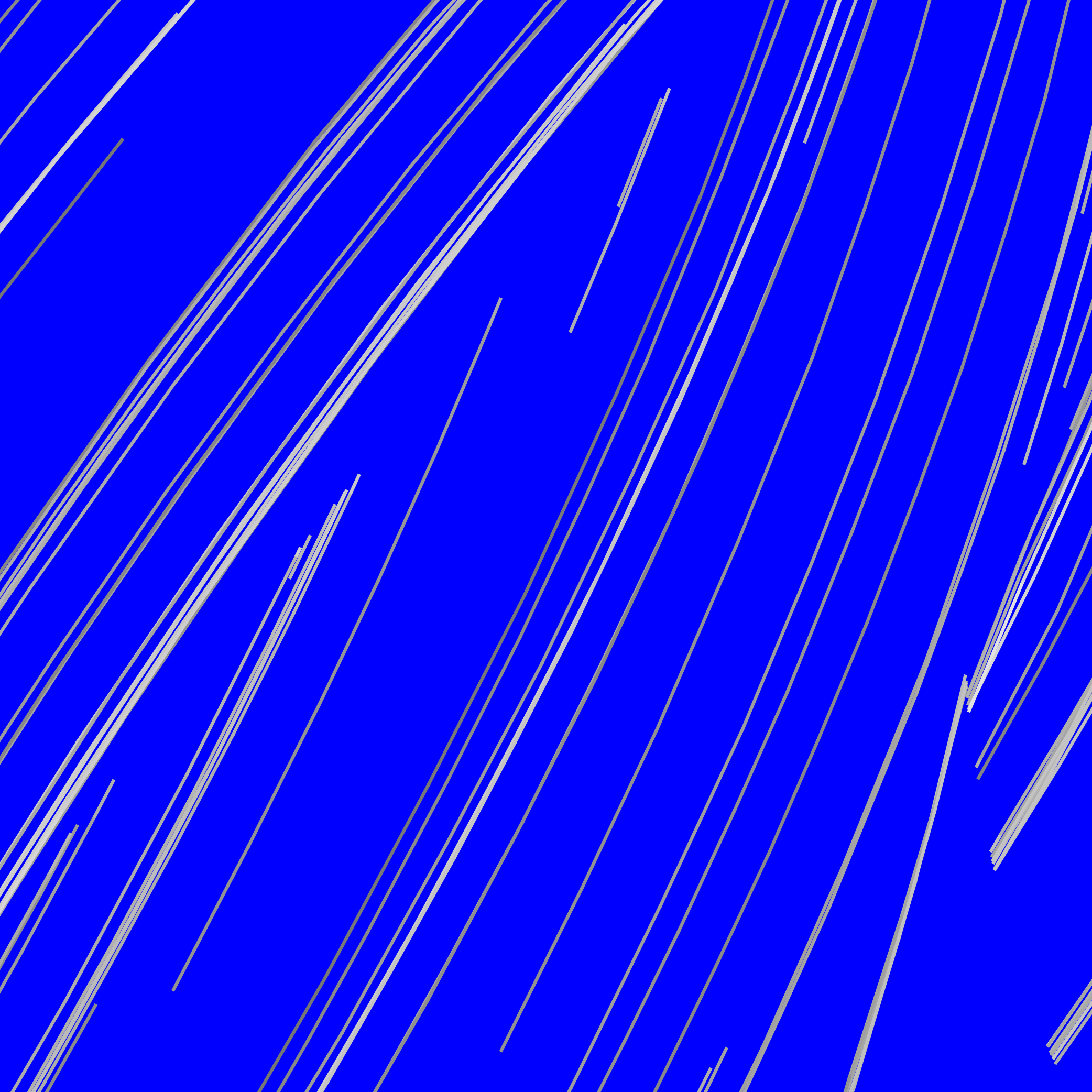}&
\includegraphics[width=0.33\textwidth]{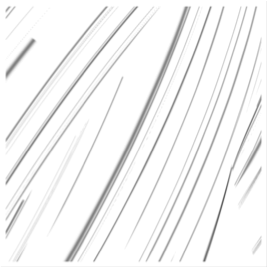}\\
(a) & (b) & (c)
\end{tabular}
\end{centering}
\caption{(a) Typical distribution of the logarithm of operational time $\ln\tau$ (black, $\ln\tau=50$, white $\ln\tau=5$) representing the backbone of the lamellar concentration field, with detail (dashed box) shown in (b). The associated scalar field $c(\mathbf{x})$ for $Pe=10^8$ shown in (c) is calculated via the diffusive strip method~\citep{Meunier:2010aa}.}\label{fig:nummethod}
\end{figure}

 To solve~\eqref{ADEstrip}, we define the reduced coordinate $\eta_0$ and operational time $\tau(t)$~\cite[][]{Ranz79} as 
\begin{align}
\label{etatau}
\eta_0(t) = \eta \rho(t), && \tau(t) = \int\limits_0^t d
t^\prime \rho(t^\prime)^{2}, 
\end{align}
where $\rho(t) = {l(t)}/{l(0)}$. This is equivalent to a transformation into the characteristic system
of~\eqref{ADEstrip}. Using this transformation,~\eqref{ADEstrip} simplifies to the
diffusion equation
\begin{align}
\frac{\partial g(\eta_0,\tau)}{\partial \tau} = 
D_m \frac{\partial^2 g(\eta_0,\tau)}{\partial \eta_0^2},
\end{align}
whose solution for the Gaussian initial condition~\eqref{c0} is given by $g(\eta_0,\tau) = \exp[-\eta_0^2/2(\sigma_0^2 + 2 D_m \tau)]/\sqrt{2 \pi(\sigma_0^2 + 2 D_m \tau)}$. Hence the concentration profile
$c(\eta,t) = g[\eta_0(t),\tau(t)]$ across the strip is
\begin{align}
\label{sol:gauss}
c(\eta,t) = \frac{c_0}{\sqrt{2 \pi
    \left[\sigma_0^2 + 2 D_m \tau(t)\right]}}\exp\left[-\frac{\eta^2 \rho(t)^2}{2
      \sigma_0^2 + 4 D_m\tau(t)} \right],
\end{align}
and so the transverse concentration profile is completely determined by the total stretch $\rho(t)$ which acts to stretch and narrow the lamellae and the stretching history as quantified by $\tau(t)$ which acts to broaden and dilute the lamellae. As $\tau(t)$ depends explicitly upon the entire stretching history, the stretching oscillations in the two-step process (\ref{eqn:twostepCTRW}) over a couplet are captured by $\tau(t)$, and so must be appropriately quantified in the implementation of the one-step CTRW (\ref{onestepCTRW}). The maximum concentration $c_{m}(t) = c(0,t)$ is now given by
\begin{align}
c_{m}(t) = \frac{c_0}{\sqrt{2 \pi \left[\sigma_0^2 + 2 D_m \tau(t)
    \right]}}.\label{eqn:cmax_dimensional}
\end{align}
%
These results form the basis of the diffusive strip method~\citep{Meunier:2010aa} which facilitates efficient solution of the 1D lamellar ADE (\ref{ADEstrip}) based upon the advection of fluid particles across a very broad range of $Pe$. We employ this method in conjunction with the advective $\mathcal{M}$ and temporal $\mathcal{T}$ maps over a pore couplet to rapidly simulate stretching of a 2D material filament within the 3D pore-space, and diffusion is calculated as a post-processing step for various $Pe$ via (\ref{sol:gauss}) given determination of the distribution of $\rho$ and $\tau$ along the strip from the stretching history. This method is capable of accurately capturing stretching, diffusion and coalescence of the scalar field up to $n\sim 100$ pores, beyond which exponential growth of the lamellar structure (and the associated number of  representative points) is too large for feasible computation. A typical distribution of the operational time $\ln\tau$ over the cross-section of a 2D filament at longitudinal distance $n=40$ pores is shown in Figure~\ref{fig:nummethod} along with the scalar field $c(\mathbf{x},t_n)$ calculated at $Pe=10^8$.

%
We non-dimensionalise \eqref{eqn:cmax_dimensional} by introducing the variables 
\begin{align}
c^\prime_{m}(t) = \frac{c_m(t)}{c_{m}(0)}, && t^\prime = \frac{t}{\Delta
  t_a} , && \tau^\prime = \frac{\tau}{\Delta t_a}, 
\end{align}
with the characteristic advection time $\Delta t_a = \ell / v_0$;
$\ell$ is the length of a pore and $v_0$ the mean pore velocity. In
the following, we drop the primes for simplicity of notation. The
maximum concentration is then
\begin{align}
c_{m}(t) = \frac{1}{\sqrt{1 + \tau(t) / Pe}},
\label{eqn:cmax}
\end{align}
where the Pecl\'{e}t number	 is defined as $Pe = \sigma_0^2 v_0 / (2 D_m \ell)$. Dispersion across a material filament is dependent upon $Pe$ and the entire stretching
history encoded via the operational time $\tau(t)$.  

\begin{figure}
\begin{centering}
\begin{tabular}{c c}
\includegraphics[width=0.42\columnwidth]{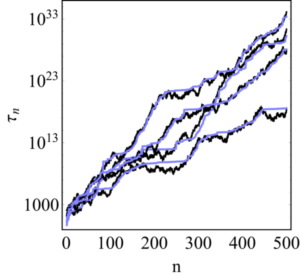}&
\includegraphics[width=0.58\textwidth]{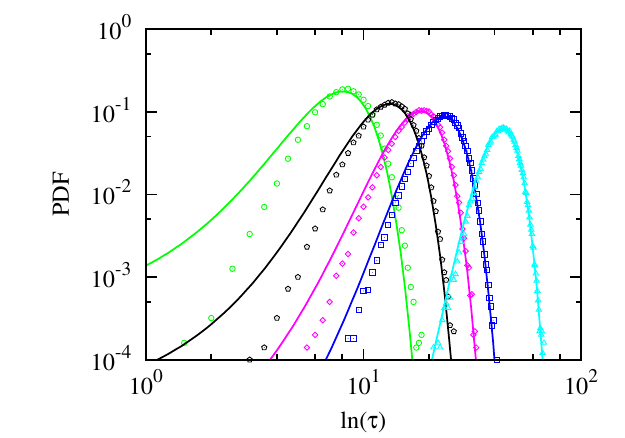}\\
(a) & (b)
\end{tabular}
\end{centering}
\caption{(a) Comparison of direct computation (black) and analytic approximation (\ref{lntau}) (blue) for operational time $\tau_n$ with pore number $n$ over four different realizations of the CTRW model. (b) Comparison of the PDF of
  $\ln\tau$ based upon the analytic approximation
  (\ref{eqn:gauss_approx_tau}) (solid lines), the PDF of $\ln\tau$
  from the CTRW model for (circles) $n = 20$, (pentagons) $n = 40$,
  (diamonds) n = 60, (squares) $n = 80$, (triangles) $n = 160$.
}\label{fig:history}
\end{figure}

The characterization of the concentration profile at location across
the strip at the downstream position $z_n$ is conditional on the
quantification of the operational time $\tau(t_n)$ at time $t_n$. As
expressed by~\eqref{etatau}, operational time depends on the deformation
history through $\rho(t)$. In order to obtain a closed form
expression for $\tau(t_n)$, we approximate the evolution of $\rho(t)$
between $t = 0$ and $t = t_n$ in terms of the average stretching rate
$\Gamma_n = {\ln(\rho_n)}/{t_n}$ after $n$ steps as
\begin{align}
\rho(t) \approx \exp\left(\Gamma_n t\right). 
\label{eqn:l_history}
\end{align}
From (\ref{eqn:l_history}), the operational time $\tau(t_n)$ may then be approximated as
\begin{equation}
\tau_n  \equiv \tau(t_n) \approx \frac{t_n
  \rho_n^2}{2 \ln(\rho_n)}. \label{eqn:tau_approx} 
\end{equation}
This expression can be further simplified by noticing that 
the PDF~\eqref{eq:pdf_rho} of $\ln(\rho_n)$ is Gaussian, with mean $n \Lambda_\infty/2$ and variance $n \sigma^2/2$. Thus, we may approximate the denominator in (\ref{eqn:tau_approx}) by its mean value $2 \ln(\rho_n) \approx n \Lambda_\infty$,
\begin{align}
\label{lntau}
\tau_n \approx \frac{t_n \rho_n^2}{n \Lambda_\infty},
\end{align}
which becomes sharper with increasing $n$ as shown in Figure~\ref{fig:history}(a). In essence, (\ref{lntau}) solves the CTRW for molecular diffusion. Oscillations associated with the two-step stretching process (\ref{eqn:twostepCTRW}) appear to play a minor role in the overall evolution of the operational time $\tau(t)$, hence $\tau_n$ is well-approximated by the one-step stretching process (\ref{onestepCTRW}) characterized by the mean stretching rate $\Lambda_\infty$. Thus, whilst diffusion is in principle dependent upon the entire stretching history, the operational time $\tau_n$ which controls the diffusion process is predominantly governed by the arrival time $t_n$ of the material strip at position $z_n$ and the elongation $\rho_n$.

The PDF of $\xi_n = \ln(\tau_n) = \ln(t_n) + 2 \ln(\rho_n) - \ln(n \Lambda_\infty)$ can be estimated by 
approximating $\ln(t_n)$ by its average 
\begin{align}
\langle \ln t_n \rangle \approx \ln\left[ \frac{n}{a} \left( \ln 2 +
    \gamma + b \right) + n \left(\ln n + 1 - \gamma \right) \right], 
\end{align}
where $\gamma$ is the Euler-Mascheroni constant and $a = 0.7413$ and $b = 0.0064$, see Appendix~\ref{App:Levy}. Since $\ln(\rho_n)$ is Gaussian distributed with the mean $n \Lambda_\infty/2$ and variance $n
\sigma^2/2$, we obtain for the PDF of $\xi_n = \ln(\tau_n)$ the Gaussian
\begin{align}
\label{eqn:gauss_approx_tau}
p_{\ln \tau}(\xi,n) = \frac{\exp\left(- \frac{[\xi - \mu_{\ln \tau}(n)]^2}{2
      \sigma{\ln \tau}^2(n)} \right)}{\sqrt{2 \pi \sigma_{\ln \tau}^2(n)}}
\end{align}
with the mean and variance
\begin{align}
\label{tau_params}
\mu_{\ln \tau}(n) = \langle \ln(t_n) \rangle + n \Lambda_\infty - \ln(n
\Lambda_\infty), && \sigma_{\ln \tau}^2(n) = 2 n \sigma^2.
\end{align}
Figure~\ref{fig:history}(b) compares the PDF of $\xi_n$ obtained from evaluating $\tau(t_n)$ according
to~\eqref{etatau} with $\rho(t_n)$ given by the CTRW~\eqref{eqn:twostepCTRW} to
the approximation~\eqref{eqn:gauss_approx_tau} (with
\eqref{tau_params}), the accuracy of which increases with $n$ due to the central limit theorem.
%
\begin{figure}
\begin{centering}
\includegraphics[width=0.6\textwidth]{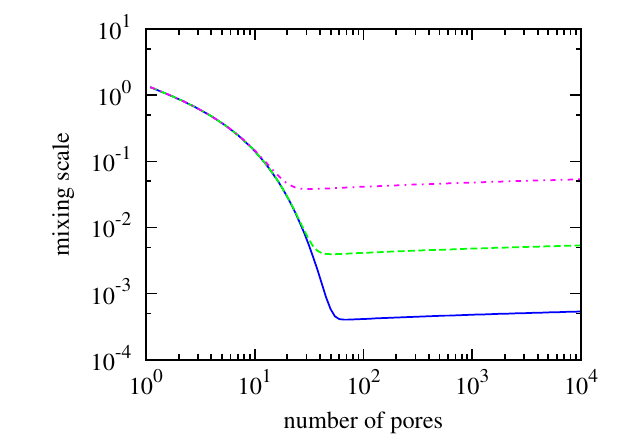}
\caption{Evolution of the average mixing scale~\eqref{epsm} normalized
  by $\sigma_0$ as a number of
  pores along the mean flow direction for (dash-dotted) $Pe = 10^4$,
  (dashed) $Pe = 10^6$ and (solid) $Pe = 10^8$.  
}\label{fig:mixingscale}
\end{centering}
\end{figure}

\section{Chaotic Mixing in 3D Random Porous Media\label{sec:diffmix}}

The approximation (\ref{eqn:gauss_approx_tau}) for $\tau_n$ provides an accurate solution to the two-step CTRW (\ref{eqn:twostepCTRW}), which along with the distribution (\ref{eq:pdf_rho}) for $\rho_n$ fully quantifies evolution of the concentration distribution, mixing and dilution in the 3D open porous network. In the following we apply this solution to determine evolution of the mixing scale, concentration PDF,  maximum concentration, scalar variance and the onset of coalescence in the 3D random porous network.

\subsection{Mixing Scale and Onset of Coalescence}

The mixing scale $\epsilon_m$ characterizes the distribution of lamellae widths at position $z_n = n$ as
\begin{align}
\label{em}
\epsilon_m(n) = \frac{\int_{-\infty}^\infty d \eta |\eta| c(\eta, t_n)}{\int_{-\infty}^\infty d
  \eta  c(\eta,t_n)} = \sigma_0 \frac{
  \sqrt{2(1 + \tau_n / Pe)}}{\pi \rho_n},
\end{align}
%
%
%
the average of which is well approximated by substitution of the approximation~\eqref{lntau} for $\tau_n$ as
\begin{align}
\langle \epsilon_m(n) \rangle \approx \sigma_0 \left\langle \sqrt{\frac{2
    (\langle t_n \rangle + Pe\,\rho_n^{-2} )}{\pi n
   Pe \Lambda_\infty}} \right\rangle,  
\end{align}
Note that whilst $\langle t_n \rangle$ is strictly infinite, the above average is dominated by the bulk of $p_n(t)$, which, as outlined in Appendix~\ref{App:Levy} can be well
approximated by the Moyal distribution. Thus the average $\langle t_n \rangle$ is understood to be the average of the equivalent Moyal distribution given by~\eqref{apptav}. In order to perform the average over $\rho_n$, we use a saddle point approximation which yields
\begin{align}
\langle \epsilon_m(n) \rangle \approx \sigma_0 \sqrt{\frac{2 \ln n}{\pi
   Pe \Lambda_\infty}} \sqrt{1 + \frac{\ln 2 +
    \gamma + b}{a \ln n} + \frac{1 - \gamma}{\ln n} + \frac{Pe
  \exp(-\Lambda_\infty n)}{n \ln n}}.  
\label{epsm}
\end{align}
This is a remarkable result because although fluid stretching due to pore-scale chaos grows fluid elements exponentially with longitudinal pore number $n$, the mixing scale does not converge to a constant Batchelor scale with increasing $n$, but rather increases asymptotically as
\begin{align}
\langle \epsilon_m(n) \rangle \sim \sigma_0 \sqrt{\frac{2 \ln n}{\pi
   Pe \Lambda_\infty}}.  \label{eqn:meanmixscale}
\end{align}
This can be traced back to the broad distribution of arrival times between the couplets arising from the no-slip condition, which renders a distribution of stretching rates of variable strength. Note also, that the  characteristic waiting time between stretching events increases with increasing number of couplets, and thus, the stretching rate decreases. This is a characteristic of the Pareto transit time distribution.  Figure~\ref{fig:mixingscale} illustrates the evolution of the average mixing scale given by~\eqref{epsm} for different Pecl\'{e}t numbers. It assumes a minimum value at a characteristic pore number $n_c \approx \ln Pe/\Lambda_\infty$, at which point diffusive expansion and compression equilibrate. 
%
\begin{figure}
\begin{centering}
\includegraphics[width=0.45\textwidth]{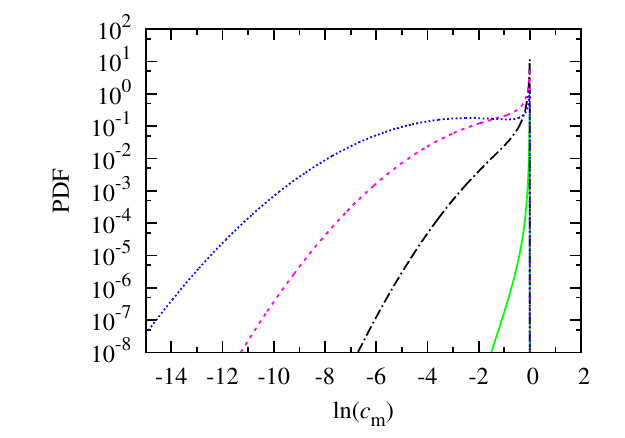}
\includegraphics[width=0.45\textwidth]{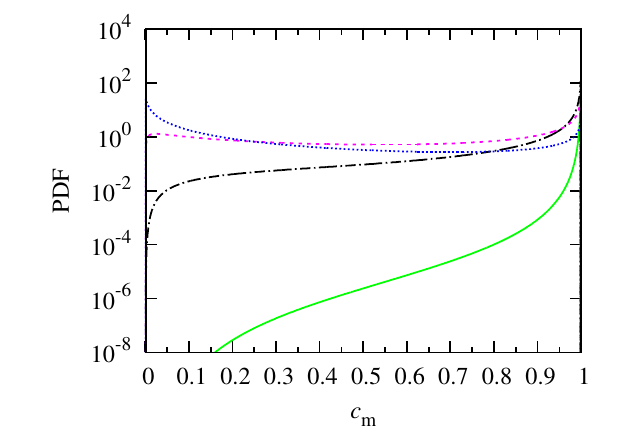}
\end{centering}
\caption{PDF~\eqref{pdfchi} of (left panel) $\ln c_m$ and (right
  panel) $c_m$ at downstream positions of (solid) $n = 20$,
  (dash-dotted) $n = 40$, (long dashed) $n = 60$ and (short dashed) $n = 80$ for
  $Pe = 10^8$. 
}\label{fig:cmpdf}
\end{figure}

Whilst the 1D lamellar ADE (\ref{ADEstrip}) is only valid as long as lamellae are non-interacting, methods are available \citep{Duplat::Villermaux::JFM::2008, Villermaux_12, LeBorgne2013} to predict evolution of the scalar concentration field in the presence of coalescence. Whilst such prediction is beyond the scope of this paper, it is important to determine the onset of coalescence and hence the envelope of validity of the model. The onset of coalscence occurs when the mixing scale $\langle\epsilon_m(n)\rangle$ exceeds the average spacing
$\epsilon(n)=\pi R^2/l(n)$ between lamellae of length $l(n)=l_0\exp(\lambda_\infty n)$ with an initial length $l_0$ in pores of average radius $R$. From (\ref{eqn:meanmixscale}), the lamellae are non-interacting up to pore number
\begin{equation}
n+\frac{1}{2\lambda_\infty}\ln \ln n\leqslant \frac{1}{\lambda_\infty} \ln\left(\frac{\pi R^2}{\l_0\sigma_0}\sqrt{\frac{\pi Pe\Lambda_\infty}{2}}\right),\label{eqn:coalescence}
\end{equation}
where the linear contribution on the LHS of (\ref{eqn:coalescence}) is due to exponential stretching of the lamellae, and the weaker nonlinear term is due to evolution of the mixing scale $\langle\epsilon_m(n)\rangle$.

\subsection{PDF of Maximum Concentration}
%
The dimensionless maximum concentration $c_m(t)$ as a function of pore number $n$, $c_m(n) = c_m(t_n)$
is given by~\eqref{eqn:cmax} as
\begin{align}
 c_{m}(n) = \frac{1}{\sqrt{1 + \tau_n/Pe}}. 
\label{eqn:cmax:3}
\end{align}
In order to develop an analytic expression for the PDF of $c_m(n) = c_m(t_n)$, we express~\eqref{eqn:cmax:3} as a function of $\xi_n = \ln(\tau_n)$, which is distributed according to~\eqref{eqn:gauss_approx_tau} as
\begin{align}
\xi_n = \ln\left(Pe \left[c_{m}(n)^{-2} - 1\right]\right),
\label{eqn:cmax:4}
\end{align}
hence the PDF $p_m(c_m,n)$ of $c_{m}(n)$ is then
\begin{align}
p_m(c_m,n) = \frac{2}{c_m (1 - c_m^2)} p_{\ln \tau}\left[\ln\left(Pe
    \left[c_{m}^{-2} - 1\right]\right) \right],
\end{align}
and the PDF of $\zeta_n = \ln[c_m(n)]$ is given by
\begin{align}
\label{pdfchi}
p_{\ln c_m}(\zeta,n) = \frac{2 }{1 - \exp(2 \zeta)} p_{\ln \tau}\left[\ln\left(Pe
    \left[\exp(-2 \zeta) - 1\right]\right) \right].
\end{align}
Notice that for $|\zeta| \gg 1$ the PDF of $\zeta$ is well approximated by a Gaussian distribution, as per Figure~\ref{fig:cmpdf} which shows the PDF of $\ln
c_m$ for $Pe=10^8$ at longitudinal distances of $n = 20,\;40,\;60$ and $80$ pores. For small $n\leq 40$, the PDF is sharply peaked about the initial concentration of $c_m = 1$, but for $n > 40$ a peak starts forming away from $c_m=1$ and the bulk of the probability weight moves away from this initial concentration. This behaviour is reflected by
evolution of the mixing scale with $n$ (illustrated in Figure~\ref{fig:mixingscale}), which assumes its minimum at $n_c \approx \ln Pe/\Lambda_\infty$, which for the $Pe = 10^8$
chosen here is $n_c \approx 40$. Thus, once the mixing scale assumes its minimum value, dilution increases markedly due to increased diffusive mass transfer.

\begin{figure}
\begin{centering}
\includegraphics[width=0.6\textwidth]{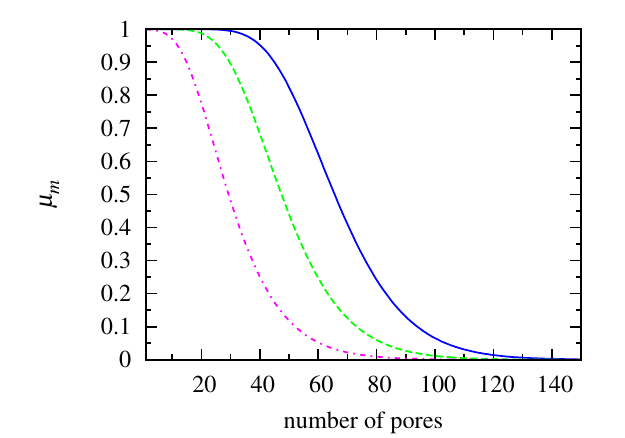}
\caption{Evolution of the average maximum concentration as a number of
  pores along the mean flow direction for (dash-dotted) $Pe = 10^4$,
  (dashed) $Pe = 10^6$ and (solid) $Pe = 10^8$.  
}\label{fig:cm}
\end{centering}
\end{figure}

The average maximum concentration $\langle c_m(n) \rangle$ shown in Figure~\ref{fig:cm} evolves in a similar manner; for $n < n_c$ the average maximum concentration $\langle c_m(n)\rangle$ is essentially equal to the initial concentration of $1$. For $n \gg n_c$, it decays exponentially rapidly according to 
\begin{align}
\label{cmapprox}
\langle c_m(n) \rangle \approx \exp\left[\frac{\sigma_{\ln \tau}^2(n)}{8} -
  \frac{\mu_{\ln \tau}(n) - \ln Pe}{2}\right],  
\end{align}
where $\mu_{\ln \tau}$ and $\sigma^2_{\ln \tau}$ are given by~\eqref{tau_params}. Hence exponential fluid stretching due to chaotic advection in 3D porous media generates exponential dilution.

Conversely, fluid deformation in 2D porous media is limited to algebraic
stretching as a consequence of the Poincar\'{e}-Bendixson theorem. In such media \citet{LB2015} and\citet{Dentz:2015aa} find explicitly
\begin{align}
\label{params_2d}
\mu_{\ln \rho,\text{2D}}(n) = \alpha \ln(n+1) + \sigma^2_{\ln \rho,\text{2D}}(n), && \sigma^2_{\ln
  \rho,\text{2D}}(n) = \beta \ln(n +1), 
\end{align}
where both $\alpha,\beta \in[1/2,2]$ . The mean and variance
of the Gaussian $\ln \tau_n$ PDF~\eqref{eqn:gauss_approx_tau} for 2D porous media are then 
\begin{align}
\label{tau_params_2d}
\mu_{\ln \tau,\text{2D}}(n) = \langle \ln(t_n) \rangle + 2 \mu_{\ln \rho,\text{2D}}(n) -
\ln[2 \mu_{\ln \rho,\text{2D}} (n)], && \sigma_{\ln \tau,\text{2D}}^2(n) = 4 \sigma^2_{\ln
\rho,\text{2D}}(n),
\end{align}
and from~\eqref{cmapprox}, we obtain the asympotic algebraic decay of average maximum concentration as
$\langle c_{m,\text{2D}}(n) \rangle \propto n^{-\alpha - 1/2 +
  \frac{\beta}{2}}$. Hence there exists a qualitative difference in fluid mixing between 2D and 3D porous media: in 2D media fluid stretching is constrained to be algebraic, leading to algebraic dilution, whereas exponential stretching is inherent to 3D porous media, yielding dilution which scales exponentially with longitudinal distance.
 
\begin{figure}
\begin{centering}
\includegraphics[width=0.6\textwidth]{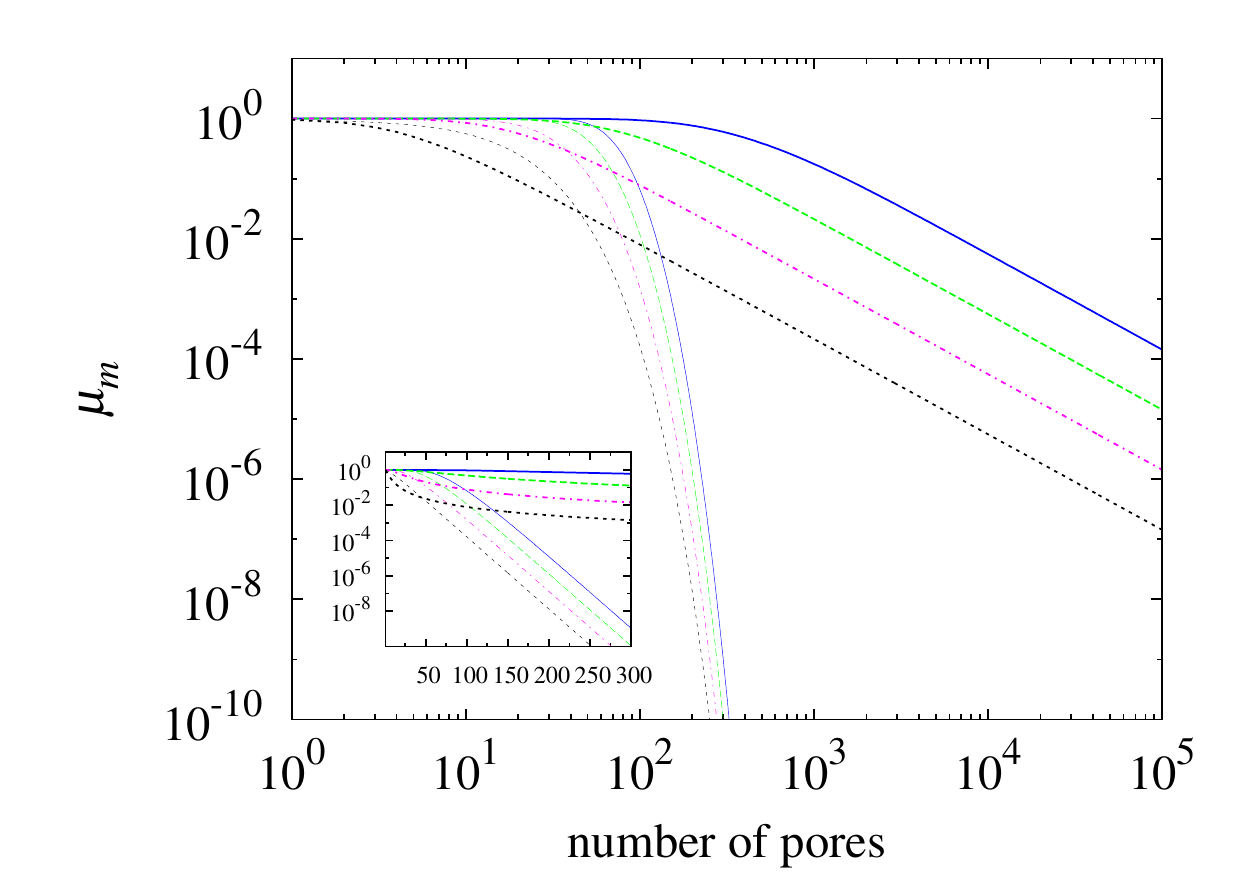}
\caption{Evolution of the average maximum concentration as a function of the number of
  pores along the mean flow direction for (short-dotted) $Pe = 10^2$  (dash-dotted) $Pe = 10^4$,
  (long-dashed) $Pe = 10^6$ and (solid) $Pe = 10^8$. The thin lines
  indicate the $3D$ pore mixing model, the thick lines, the $2D$ pore
  mixing model~\eqref{params_2d}--\eqref{tau_params_2d} with $\alpha = 1$ and $\beta =
  1/5$. The inset illustrates the same plot in a semi-logarithmic
  scale. 
}\label{fig:cmcomp}
\end{centering}
\end{figure}

\subsection{PDF of Concentration}

\begin{figure}
\begin{centering}
\includegraphics[width=0.45\textwidth]{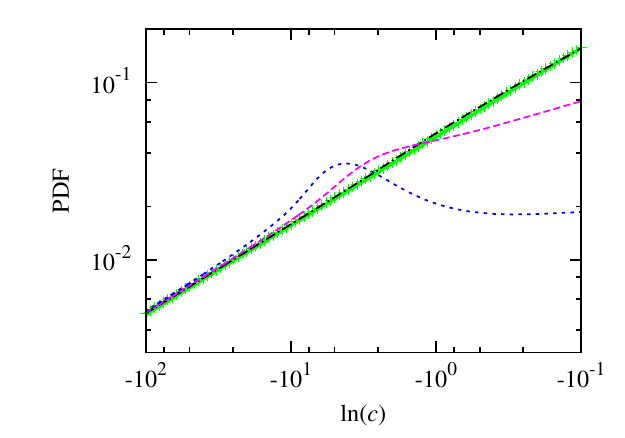}
\includegraphics[width=0.45\textwidth]{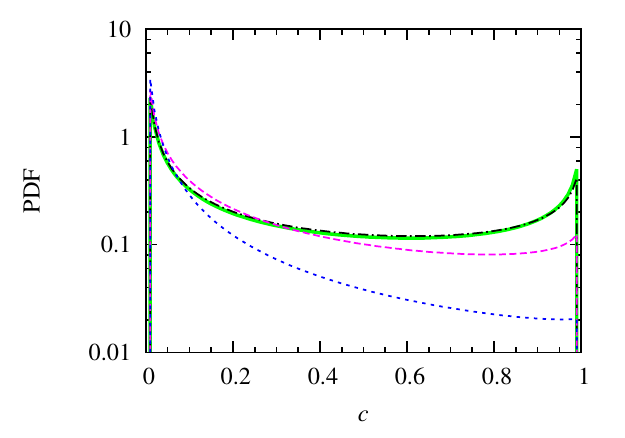}
\end{centering}
\caption{PDF of (left panel) $\ln c$ and (right panel) $c$ at downstream positions of (solid) $n = 20$,
  (dash-dotted) $n = 40$, (long dashed) $n = 60$ and (short dashed) $n = 80$
  pores for $Pe = 10^8$.}\label{fig:cpdf}
\end{figure}

We derive the concentration PDF, mean and variance within the plume as a function of the longitudinal pore number $n$. As such, this PDF is defined with respect to a support volume that is a subset of the fluid domain which excludes negligible concentrations beyond a minimum cutoff value $\epsilon$. To determine this concentration PDF, we note that the concentration PDF across a single lamella for a given maximum
concentration $c_m$ is obtained from~\eqref{sol:gauss} through spatial mapping as 
%
\begin{align}
\label{pcm}
p(c|c_m) = \frac{1}{2 c \sqrt{\ln(c_m/\epsilon) \ln(c_m/c)}},
\end{align}
where the concentration range under consideration is $[\epsilon,c_m]$ with $\epsilon$ a minimum concentration. The global concentration PDF is then
\begin{align}
p(c,n) = \int\limits_{c}^\infty d c_m p(c|c_m) p_m(c_m,n), 
\end{align}
which, using ~\eqref{pdfchi} and \eqref{pcm}, may be expressed in terms of the PDF~\eqref{eqn:gauss_approx_tau} of $\ln \tau$ as 
\begin{align}
\label{eqn:cpdf}
p(c,n) = \int\limits_{\ln c}^\infty d z \frac{1}{1 - \exp(2 z)}
\frac{p_{\ln \tau}\left[\ln\left(Pe
    \left[\exp(2 z) - 1\right],n\right) \right]}{c\sqrt{(z-\ln \epsilon)(z -\ln c)}},
\end{align}
which may be simplified via a saddle point approximation for $\ln c < \mu_c = \ln
\langle c_m \rangle$ to 
\begin{align}
\label{pca}
p(c,n) \approx \frac{1}{2 c\sqrt{(\mu_c - \ln \epsilon)(\mu_c -\ln c)}}. 
\end{align}
Figure~\ref{fig:cpdf} shows the PDFs of $\ln c$ and $c$ from~\eqref{pca} for $Pe=10^8$ as a function of longitudinal pore number $n$ for $n = 20,\;40,\;60$ and $80$ pores. The concentration PDF follows the same trend as that for maximum concentrations in that the PDFs for $n = 20$ and $40$ are almost indistinguishable due to limited dilution for $n \leq n_c$ (where $n_c \approx 40$ for $Pe = 10^8$), whereas for significant dilution arises for $n>n_c$ after the mixing scale reaches its minimum.

\subsection{Concentration Mean and Variance}



The concentration PDF determined in the previous section is obtained by sampling the concentration 
values $c(x,y,n)$ in space over the concentration support  $A_c$ which is defined as $\{\vx|c(x,y,n) > \epsilon\}$ where $\epsilon$ is the minimum concentration threshold and $A_c$ is a subset of the pore cross-sectional area $A$. Thus, the concentration PDF may be written as 
\begin{align}
p(c,n) = \frac{1}{A_c(n)} \int\limits_{A_c(n)} d^2 \mathbf x \delta[c - c(x,y,n)],
\end{align}
and the areal concentration support $A_c(n)$ is given by 
\begin{align}
A_c(n) = \int d^2\mathbf x H[c(x,y,n) - \epsilon]. 
\end{align}
%
%
%
Whilst the concentration support $A_c(n)$ quantifies mixing within the plume, for some applications it is useful to quantify mixing and dilution over the entire fluid cross-sectional area as support so to quantify global dispersion throughout the pore-space. The decay of such fluid-support spatial variance describes mixing and dilution of the concentration field toward a homogeneous state across the entire pore volume. Conversely, the concentration support measure describes dilution within an evolving plume which deliberately avoids reference to the fluid bulk, hence avoiding the singularity at $c=0$ in the concentration PDF. We denote the $k$-th moment of the spatial concentration PDF under the concentration and fluid supports respectively as
 \begin{align}
\langle c(n)^k\rangle &\equiv\frac{1}{A_c}\int\limits_{A_c} d \mathbf x \, c(x,y,n)^k,\label{eqn:concsup}\\
\overline{c(n)^k} & \equiv\frac{1}{A} \int\limits_{A} d \mathbf x\, c(x,y,n)^k \approx \frac{1}{A} \int\limits_{A_c} d \mathbf x\, c(x,y,n)^k,\label{eqn:fluidsup}
\end{align}
and so these moments are related as $A\,\overline{c(n)^k}\approx A_c\langle c(n)^k\rangle$ for $\epsilon\ll c_m$. To derive the concentration mean and variance throughout the pore network under both of these measures, we consider the evolution of a single lamella which originates at the injection plane ($z=n=0$) as a strip of length $l_0$ with transverse concentration distribution $c(\eta,0)$ given by (\ref{c0}). We pose the spatial distribution of the concentration field after $n$ pores in terms of the material coordinates $(\eta,\zeta)$, where $\zeta$ is the Lagrangian material coordinate along the 1D lamella backbone (shown in Figure~\ref{fig:nummethod}(b)), the domain of which $\zeta\in[0,l_0]$ references the injected lamella at $z=n=0$.

Note that the lamella segments distributed throughout a given pore (at fixed $z$) comprise of contributions from different lamellae injected at $z=n=0$. Whilst the material coordinate $\zeta$ refers to an individual lamella sheet, 
in the present context we interpret $\zeta$ to denote the material coordinate along all of the lamella segments in a given pore, the union of which also span $\zeta\in[0,l_0]$ due to homogeneity of the injection protocol. Whilst this simplification does not extend to inhomogeneous injection protocols such as a point or line source, the homogeneous injection results may readily be generalised to inhomogeneous protocols as per $\S$\ref{sec:point_inject}.

 At pore number $n$ the advection time $t_n$, deformation $\rho$ and operational time $\tau$ all vary with the material coordinate $\zeta$ along the lamella, and so we may parameterise these quantities in terms of $\zeta$. Hence the 2D spatial Gaussian concentration distribution (\ref{ADEstrip})  over the entire lamella may be expressed in material coordinates $(\eta,\zeta)$ as
\begin{equation}
c(\eta,\zeta)=\frac{c_0}{\sqrt{2\pi\sigma_{\eta}^2(\zeta)}}\exp\left[-\frac{\eta^2 \rho(\zeta)^2}{2\sigma_\eta^2(\zeta)}\right],\label{eqn:spatial_conc}
\end{equation}
where the concentration variance $\sigma_\eta^2(\zeta)=\sigma_0^2+2D_m\tau(\zeta)$.
The concentration support $A_{c}(n)$ is quantified by the area of the concentration field for which $c(\eta,\zeta,n)\geqslant\epsilon$, which corresponds to the cutoff length $\eta_\epsilon(\zeta)$ from the lamellar backbone $\eta=0$ as
\begin{equation}
\eta_\epsilon(\zeta) \equiv \frac{\sigma_\eta(\zeta)}{\rho(\zeta)}\sqrt{2\ln\left[\frac{c_m(\zeta,n)}{\epsilon}\right]},
\end{equation}
%
%
%
%
and so the concentration mean $\overline{c(n)}$ under the fluid support for $\epsilon\ll c_m(\zeta,n)$ is then
\begin{equation}
 \begin{split}
\overline{c(n)}
\approx\frac{1}{A} \int_0^{l_0} d\zeta\,\rho(\zeta) \int_{-\eta_\epsilon(\zeta)}^{\eta_\epsilon(\zeta)} d\eta\,c(\eta,\zeta)
=\frac{c_0 l_0}{A} \left\langle \text{erf}\left[\sqrt{\ln\frac{c_m(\zeta,n)}{\epsilon}}\right] \right\rangle
\approx \frac{c_0 l_0}{A}.\label{eqn:avconc}
 \end{split}
 \end{equation}
Note that the integration over $\zeta$ weighted by $l_0$ is equivalent to performing the average over the ensemble of the elementary lamellae 
due to the ergodicity of the system as discussed above. As a consequence of~\eqref{eqn:avconc} and~\eqref{eqn:concsup}, the concentration support area $A_c(n)$ evolves as
\begin{equation}
A_c(n)\approx\frac{c_0 l_0}{\langle c(n)\rangle},
\label{eqn:Ac}
\end{equation}
%
where $A_c(n)$ increases and $\langle c(n)\rangle$ decreases with $n$ as per the concentration support PDF (\ref{eqn:cpdf}). 
The concentration support $A_c(n)$ can be determined by integration along all the lamella segments as 
\begin{align}
A_c(n) = \int\limits_{0}^{l_0}  d \zeta \rho(\zeta) \int\limits_{-\eta_\epsilon(\zeta)}^{\eta_\epsilon(\zeta)}  d\eta = 
\sqrt{\frac{\pi}{2}}
\left\langle \frac{c_0 l_0}{c_m(\zeta,n)} \sqrt{2\ln\left[\frac{c_m(\zeta,n)}{\epsilon}\right]} \right\rangle. 
\end{align}
Thus, it increases approximately as $\langle c_m(n)^{-1}\rangle$. 

The difference between the fluid- and concentration-support measures is clearly reflected by the behaviour of the means under these respective measures; under the fluid-support the mean concentration is constant due to conservation, whereas under the concentration support the mean concentration within the plume is decreasing due to plume spreading and dilution. Whilst the result (\ref{eqn:avconc}) holds for all $n$ due to conservation of mass, the derivation above, (\ref{eqn:Ac}) and the PDF (\ref{eqn:cpdf}) is only valid in the pre-coalescence regime where $A_c$ does not overlap.
%
%
To calculate the fluid and concentration support concentration variances 
\begin{align}
\sigma_{\overline{c}}^2(n)&\equiv\overline{c(n)^2}-\overline{c(n)}^2,\\
\sigma_{\langle c\rangle}^2(n)&\equiv\langle c(n)^2\rangle-\langle c(n)\rangle^2,
\end{align}
under $\epsilon \ll c_m(\zeta)$ we find
\begin{equation}
\begin{split}
\int\limits_{-\eta_\epsilon(\zeta)}^{\eta_\epsilon(\zeta)} d \eta \,c(\eta,\zeta)^2=\frac{c_0^2}{2 \sqrt{\pi}\rho(\zeta)\sigma_\eta(\zeta)}\text{erf}\left[\sqrt{\ln\frac{c_m(\zeta,n)}{\epsilon}}\right]\approx\frac{c_0^2}{2\sqrt{\pi}\rho(\zeta)\sigma_\eta(\zeta)},
\end{split}
\end{equation}
and so the second concentration moment $\overline{c(n)^2}$ is then
\begin{equation}
\overline{c(n)^2}\approx\frac{\overline{c(n)}}{\sqrt{2}} \frac{1}{l_0} \int_0^{l_0}\frac{c_0}{\sqrt{2 \pi \sigma_\eta(\zeta)^2}}d\zeta =
\frac{\overline{c(n)} \langle c_m(n) \rangle}{\sqrt{2}} 
\end{equation}
where, due to stationarity between pores at fixed $n$, the average of $1/\sigma_\eta(\zeta)$ along the material coordinate $\zeta$ is equivalent to the ensemble average $\langle 1/\sigma_\eta \rangle$. As the maximum concentration $c_m=c_0/\sqrt{2\pi\sigma_\eta^2}$, then the concentration variances may be expressed directly in terms of the average maximum concentration $\langle c_m\rangle$ as
\begin{align}
\sigma_{\overline{c}}^2(n) &= \overline{c(n)}^2 \left(\frac{ \langle c_m(n) \rangle}{\overline{c(n)}\sqrt{2} }-1 \right)\label{eqn:variance},\\
\sigma_{\langle c\rangle}^2(n) &=\langle c(n) \rangle^2 \left(\frac{ \langle c_m(n) \rangle}{\langle c(n)\rangle\sqrt{2} }-1 \right).\label{eqn:concvariance}
\end{align}
Note that the mean maximum concentration $\langle c_m\rangle$ is neither a fluid- or concentration- support measure but rather is averaged with respect to the 1D manifold comprised of the lamellar segments in the pore cross-section. Whilst (\ref{eqn:variance}), (\ref{eqn:concvariance}) yield negative concentration variances in the homogenization limit of large $n$ (as $\langle c_m(n)\rangle\rightarrow \langle c(n) \rangle \rightarrow \overline{c(n)}$), the derived model is valid in the pre-coalescence regime, and so is not expected to capture the late time dissipation dynamics. Conversely, at earlier times when $\langle c_m(n)\rangle\gg\overline{c(n)}$,$\langle c_m(n)\rangle \gg \langle c(n)\rangle$, both measures of scalar variance evolve in direct proportion to the maximum concentration $c_m(n)$ (\ref{eqn:cmax:4}), yielding exponential scalar dissipation with pore number $n$ for 3D porous media and algebraic dissipation in 2D media. 

%
%



\subsection{Chaotic Mixing within a Point Injection Plume}\label{sec:point_inject}

Whilst the results above pertain to mixing of a concentration field which is heterogeneous at the pore-scale injected across all pores transverse to the mean flow direction, it is also instructive to consider how these results apply to dilution within a steady solvent plume arising from a continuously injected point source in the 3D random porous network, as illustrated in Figure~\ref{fig:pore_network}. For a steady plume in homogeneous porous media, the mean macroscopic concentration can be assumed to follow a Gaussian distribution in the direction transverse to the mean flow. Hence, the average lamellar length $l_p$ per pore varies with transverse radial distance $r$ and longitudinal distance $z_n=\ell n$ from the injection source point can be approximated by the 2D Gaussian distribution
\begin{equation}
l_p(n,r)=\frac{l(n)}{2 \pi n^2\sigma_{tr}^2}\exp\left(-\frac{(\varphi r)^2}{2n^2\sigma_{tr}^2}\right),\label{eqn:lamellae}
\end{equation}
where the total lamellar length is $l(n)=l_0\exp(\lambda_\infty n)$, $\varphi$ is the transverse areal porosity, and $\sigma_{tr}^2=d^2/2$ is the standard deviation associated with a pore branch or merger with centre-to-centre distance $d$. Such lateral spreading of the plume under continuous point-wise injection significantly retards the onset of coalescence, such that the condition (\ref{eqn:coalescence}) is now
\begin{equation}
n+\frac{1}{2\lambda_\infty}\ln \ln n - \frac{(\varphi r)^2}{2\lambda_\infty \sigma_{tr}^2} -\frac{2}{\lambda_\infty}\ln(n \sigma_{tr}^2) \leqslant \frac{1}{\lambda_\infty} \ln\left(\frac{R^2}{l_0\sigma_0}\sqrt{2\pi Pe\Lambda_\infty}\right).\label{eqn:plume_coalescence}
\end{equation}
Whilst measures with respect to the concentration support $A_c$ such as the mean mixing scale $\langle \epsilon_m(n)\rangle$, maximum concentration $\langle c_m(n)\rangle$, concentration PDF $p(c|n)$, concentration mean $\langle c(n)\rangle$ and variance $\sigma_{\langle c \rangle}^2(n)$ under the concentration support are the same as for point-wise or uniform injection, the concentration mean $\overline{c(n)}$ and variance $\sigma_{\overline{c}}^2(n)$ under the fluid support are markedly different. Following (\ref{eqn:avconc}) and (\ref{eqn:variance}), these quantities within the plume vary with pore number $n$ and radial distance $r$ as
\begin{align}
&\overline{c_p(n,r)} = \overline{c(n)}\frac{l_p(n,r)}{l(n)},\label{eqn:plumemean}\\
&\sigma_{\overline{c},p}^2(n,r)=\overline{c_p(n,r)}^2\left(\frac{\langle c_m(n)\rangle}{\overline{c_p(n,r)}\sqrt{2}}-1\right) \label{eqn:plumevar}.
\end{align}
Hence the concentration distribution within the plume follows the 2D Gaussian lamellar distribution (\ref{eqn:lamellae}), and the rate of scalar dissipation is given by dilution of the maximum concentration $\langle c_m(n)\rangle$ which decays exponentially with pore number $n$ in 3D random media. Note that as $\overline{c_p(n,r)}\ll\langle c\rangle$, the region of validity of (\ref{eqn:plumevar}) is significantly larger for the plume injection case. In general (\ref{eqn:plumemean}), (\ref{eqn:plumevar}) hold for any macroscopic concentration $l_p(x,y,n)/l(n)$ arising from any injection protocol at $n=0$ in both heterogeneous or homogeneous media. 

\section{Discussion\label{sec:discussion}}




The topological complexity inherent to three-dimensional porous media~\citep{Vogel:2002} imparts chaotic advection and exponential fluid stretching under steady flow conditions~\citep{Lester:2013ab}. Such complexity generates a large number density of saddle points in the skin friction field, rendering the associated stable and unstable manifolds which project into the fluid bulk two-dimensional. These 2D surfaces of minimal transverse flux~\citep{MacKay:2001,MacKay:2008} control transport and mixing, where transverse intersection generates chaotic advection dynamics and persistent exponential fluid stretching as fluid elements are advected through the pore-space. In combination with molecular diffusion, such chaotic advection significantly augments pore-scale mixing and dispersion but has received limited attention.

All porous media (both 2D and 3D, heterogenous and homogeneous) admit no-slip boundaries which impart highly heterogeneous velocity distributions. These distributions determine the frequency of stretching events under advective flow, and the no-slip condition imparts arbitrarily long waiting times between stretching events.  We show here that these two basic ingredients of pore-scale topological mixing and heterogeneous advective velocity may be integrated successfully in an analytically tractable stochastic theory that represents fluid deformation as a continuous time random walk (CTRW). The kernels of this CTRW model are quantified via pore-scale computations of fluid deformation and transport in an model 3D random open porous network, and this CTRW model is subsequently coupled to a lamellar model of diffusive mixing to provide quantitative predictions of fluid mixing and dispersion. Although algebraic deformations such as fluid shear also impact mixing, these mechanisms are asymptotically dominated by exponential stretching at the pore-scale. 

In the 3D porous network model presented here the no-slip boundary condition generates a Pareto transit time distribution $\psi(\Delta t) \sim \Delta t^{-1-\beta}$ between stretching events, with $\beta=1$ (\ref{psi}). The model can be however readily generalized to other distributions, such as measured in fluid flow simulations through porous media reconstructed from micro-tomography imaging \citep{BijeljicEA:11}.  Chaotic advection arising from steady pore-scale advection generates a log-Gaussian distribution of relative fluid elongation $\rho$ (\ref{eq:pdf_rho}) transverse to the mean flow direction, the mean of which grows exponentially with longitudinal pore number $n$ as the Lyapunov exponent $\lambda_\infty$. The interplay of exponential fluid stretching and Pareto-distributed waiting times leads to an average mixing scale $\langle\epsilon_m(n)\rangle$ (\ref{eqn:meanmixscale}) which does not converge with $n$ to a constant Batchelor scale, but rather reaches a minimum at $n=n_c\approx \ln Pe/\Lambda_\infty$ and scales asymptotically as $\sqrt{\ln n}$. Consequently, the average maximum concentration $\langle c_m(n) \rangle$ (\ref{cmapprox}) and concentration PDF (\ref{pca}) within lamellae evolve in a similar fashion, where dilution is negligible up to $n\leq n_c$, but for $n>n_c$ the lamellae then broaden and significant dilution occurs.

The impact of fluid deformation upon fluid mixing and dilution in 2D and 3D porous media is clearly illustrated in Figure~\ref{fig:cmcomp} by the different scalings for the average maximum concentration $\langle c_m(n)\rangle$ (\ref{cmapprox}). In 2D porous media, algebraic fluid stretching leads to fluid mixing which scales algebraically (\ref{params_2d}) with with pore number $n$, whereas the exponential fluid stretching associated with chaotic mixing in random 3D porous media imparts exponential mixing (\ref{cmapprox}). This behaviour is directly reflected by evolution of the spatial concentration variance under both fluid-support $\sigma_{\overline{c}}^2(n)$ (\ref{eqn:variance}) and concentration-support $\sigma_{\rangle c \langle}^2(n)$ (\ref{eqn:concvariance}) measures. These results directly quantify the impact of chaotic mixing in 3D porous media in terms of the pore-scale stretching and advection dynamics.

Whilst this model is only valid up to the coalescence of lamellae as per (\ref{eqn:coalescence}), it may be extended as per \cite{Duplat::Villermaux::JFM::2008, Villermaux_12, LeBorgne2013} to capture the coalescence regime where mixing is primarily controlled by a diffusive aggregation processes. As mixing dynamics are universally dependent on the rate of fluid deformation, we anticipate that the impact of different fluid stretching dynamics inherent to 2D and 3D random porous media shall persist throughout the coalescence regime.

Application of the CTRW model to a point source solute plume injected shows that chaotic advection again imparts exponential mixing, and the fluid-support concentration variance $\sigma_{\overline{c},p}^2(n,r)$ (\ref{eqn:plumevar}) is the same as that for the uniform case rescaled by the mean pore concentration $\overline{c(n,r)}$ (\ref{eqn:plumemean}). This result is generic to any macroscopic concentration distribution, hence exponentially accelerated mixing persists in both heterogeneous and homogeneous media. Likewise macroscopic longitudinal dispersion is also strongly augmented by chaotic advection~\citep{Lester:2014ab}. These results have significant implications for the development of macroscopic models of dispersion and dilution which recover the pore-scale mechanisms which arise from chaotic mixing in 3D porous media.

The predictions of concentration PDF and mixing rates from the stretching CTRW model compare very well with fully resolved numerical simulations over a wide range of Pecl\'{e}t numbers for the model 3D open porous network. For extension to real pore-scale architectures, the scalar deformation CTRW framework may be extended to quantify of tensorial fluid deformation via recent developments~\citep{Lester:2015aa, Dentz:2015aa} regarding the evolution of the deformation gradient tensor in 3D steady random flows. These developments facilitate statistical characterization of deformation and mixing at the pore-scale and the development of tensorial deformation CTRW models.

\section{Conclusions\label{sec:conclusions}}

Three dimensional pore networks are characterized by i) significant topological complexity inherent to all porous media and ii) highly heterogeneous velocity distributions imparted by ubiquitous no-slip conditions at pore walls, further compounded by the distribution of pore sizes. The ubiquity of these mechanisms has significant implications for the prediction and understanding of fluid mixing and macroscopic dispersion in 3D random porous media. The interplay of exponential fluid stretching and broad velocity distributions arising from the no-slip condition generates significantly accelerated mixing via the production of highly striated, lamellar concentration distributions.


We study these mechanisms in a model 3D open porous network which is homogeneous at the macroscale, and develop a CTRW model for fluid deformation and pore-scale mixing based upon high-resolution CFD simulation of Stokes flow in the network model. Predictions of this model agree very well with direct numerical simulations. Analytic estimates of the mixing dynamics show that mixing and dilution under steady state conditions is controlled by the mean and variance of the fluid stretching rates (quantified respectively by the Lyapunov exponent $\lambda_\infty$ and the variance $\sigma^2$) and the Pecl\'{e}t number $Pe$, such that the mean concentration decays exponentially with longitudinal advection in 3D random porous media, whereas mean concentration variance decays algebraically in 2D porous media. Whilst highly idealised, these basic mechanisms are universal to  3D porous media and so these results have significant implications for both modelling and understanding mixing in random media.

The developed stretching CTRW model predicts mixing rates for general fluid stretching properties and transit time distributions. Hence, we anticipate that it may be applicable to quantify mixing in a range of porous materials to decipher the role of network topology and structure upon pore scale mixing and thus upon upscaled dilution and mixing-limited reactions. The proposed framework may be extended to transient transport conditions, relevant for instance for pulse tracer injections, through the integration of longitudinal mixing processes.  


\bigskip

MD acknowledges the support of the European Research Council (ERC)
through the project MHetScale (contract no. 617511), and TLB acknowledges the support of the ERC project ReactiveFronts, and Agence Nationale de la Recherche project Subsurface Mixing and Reaction.

\appendix

\section{Distribution of Transit Times--The Landau Distribution\label{App:Levy}}
\begin{figure}
a\includegraphics[width=0.45\textwidth]{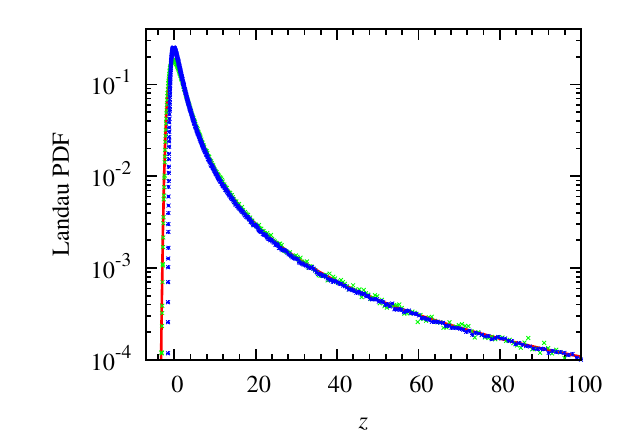}
b\includegraphics[width=0.45\textwidth]{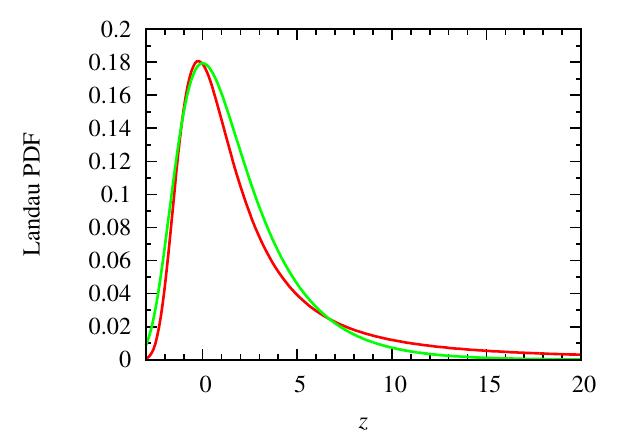}
\caption{Left:c Rescaled and shifted PDF $\hat p_n(z)$ for (blue) $n =
  10$ and (green) $n = 10^3$ obtained from random walk simulations
  for $10^6$ realizations of the stochastic process $t_n$. The red
  line indicates the Landau PDF obtained from numerical inverse
  Laplace transform of~\eqref{f1}. Right: Comparison of the (red) Landau
  PDF $f_1(z)$ defined by~\eqref{f1} and (green) the approximation~\eqref{fapprox}
by the Moyal distribution.\label{fig:landau}}
\end{figure}

The PDF $p_n(t)$ of $t_n$ given in~\eqref{eqn:twostepCTRW} can be written in
Laplace space as
\begin{align}
p_n^\ast(\lambda) = \psi^\ast(\lambda)^n. 
\end{align}
The Pareto distribution~\eqref{psi} is a Levy-stable distribution,
which means in particular that the long time behavior of $p_n(t)$ is
the same as the one of~\eqref{psi}. The Pareto
distribution~\eqref{psi} reads in dimensionless terms as
\begin{align}
\psi(\Delta t) = \frac{1}{\Delta t^2}, && \Delta t > 1. 
\end{align}
Its Laplace transform is given by 
\begin{align}
\psi^\ast(\lambda) = \exp(-\lambda) + \lambda \text{Ei}(-\lambda),
\end{align}
where $\text{Ei}(x)$ is the exponential
integral~\cite[][]{AS1972}. For small $\lambda \ll 1$, this expression
can be expanded as 
\begin{align}
\psi^\ast(\lambda) = 1 - \lambda (1 - \gamma) + \lambda
\ln(\lambda) + \dots,
\end{align}
where the dots denote subleading contributions of order $\lambda^2$,
$\gamma$ is the Euler constant. Thus, we can write $p_n^\ast(\lambda)$
for small $\lambda \ll 1$ as
\begin{align}
p_n^\ast(\lambda) = \exp\left[ - \lambda n (\ln n + 1 - \gamma) + \lambda n
\ln(\lambda n)\right]. 
\end{align}
Inverse Laplace transform of this expression, gives for $p_n(t)$ the
form
\begin{align}
\label{pnt}
p_n(t) = \frac{1}{n} f_1\left[\frac{t - n (\ln n
    + 1 - \gamma)}{n}\right],
\end{align}
where 
\begin{align}
\label{f1}
f_1(t) = \int \frac{d \lambda}{2 \pi i} \exp\left[\lambda
  \ln(\lambda)\right] \exp(-\lambda t).  
\end{align}
denotes the Landau distribution~\cite[][]{Uchaikin1999}. 
It behaves for $t \gg 1$ as $f_1(t) \approx t^{-2}$. Expression~\eqref{pnt}
describes the density $p_n(t)$in the limit of large times $t$ or large
$n$. In order to test this approximation, we performed numerical
random walk simulations for $10^6$ realizations of the random time
$t_n$. The obtained PDF $p_n(t)$ is rescaled as 
\begin{align}
\hat p_n(z) = n p_n\left[n z + n (\ln n
    + 1 - \gamma) \right],
\end{align}
In the limit $n \to \infty$, we expect $\hat p_n(z) \to
f_1(z)$. Figure~\ref{fig:landau} shows $\hat p_n(z)$ for $n = 10$ and
$10^3$ compared to~\eqref{f1}, which is obtained by numerical
inverse Laplace transform. The maximum
of~\eqref{f1} is assumed at $t_m = - \frac{1}{2} \ln \frac{\pi}{2}$,
as illustrated in Figure~\ref{fig:landau}b.

%

We consider now the average $\langle \ln t_n \rangle$, which is dominated by the bulk of the
Landau distribution $p_n(t)$. To this end, we note that the bulk of
the Landau distribution $f_1(t)$ can be approximated by the Moyal distribution
\begin{align}
f_m(x) = \frac{1}{\sqrt{2 \pi}} \exp\left[ -\frac{1}{2} \left( x +
    {\textrm e}^{-x} \right) \right]
\end{align}
as 
\begin{align}
\label{fapprox}
f_1(t) = a f_m\left[a (t-b)\right]
\end{align}
with $a = 0.7413$ and $b = 0.0064$. Thus, we may approximate $p_n(t)$ in
terms of $f_m(x)$ as
\begin{align}
\label{pnapprox}
p_n(t) \approx \frac{a}{n} f_m\left(a \left[ \frac{t - n (\ln n + 1 -
      \gamma)}{n} - b \right] \right)
\end{align}
The mean of $\ln t_n$ is then approximated by 
\begin{align}
\langle \ln t_n \rangle \approx \int\limits_{-\infty}^\infty dx \ln\left[n (\ln n + 1 -
      \gamma) + \frac{n}{a}(\langle x_m \rangle + b) + \frac{n
        x}{a}\right] f_m(\langle x_m \rangle + x),
\end{align}
where $\langle x_m = \rangle \ln 2 + \gamma$ is the mean of the Moyal
distribution. Thus, we obtain approximately for $\langle \ln t_n
\rangle$
\begin{align}
\langle \ln t_n \rangle \approx \ln\left[ \frac{n}{a} \left( \ln 2 +
    \gamma + b \right) + n \left(\ln n + 1 - \gamma \right) \right]. 
\end{align}
The mean of the $p_n(t)$ as approximated by the Moyal distribution
through~\eqref{pnapprox} is given by 
\begin{align}
\label{apptav}
\langle t_n \rangle \approx \frac{n}{a} \left( \ln 2 +
    \gamma + b \right) + n \left(\ln n + 1 - \gamma \right) . 
\end{align}
%


\end{document}